\documentclass[longauth]{aa}  
\usepackage{graphicx}
\usepackage{txfonts}
\usepackage{xcolor}
\usepackage{hyperref}
\usepackage{amsmath,siunitx}
\usepackage{threeparttable}
\newcommand{\kms}{km~s$^{-1}$}
\newcommand{\h}{H$_2$}
\newcommand{\msun}{M$_\odot$}
\newcommand{\e}{$\times10$}
\newcommand{\HII}{H\textsc{ii}}
\newcommand{\HI}{H\textsc{i}}
\newcommand{\Ha}{H$\alpha$}

\newcommand{\CO}{$^{12}$CO}
\newcommand{\Tpeak}{$T_\text{peak}$}

\begin{document} 

   \title{Quantifying the energetics of molecular superbubbles in PHANGS galaxies}

   \author{E. J. Watkins\inst{1}
          \and
          K. Kreckel\inst{1}
          \and
          B. Groves\inst{2}
          \and
          S. C. O. Glover\inst{3}
          \and
          B. C. Whitmore\inst{4}
          \and
          A. K. Leroy\inst{5,6}
          \and
          E. Schinnerer\inst{7}
          \and
          S. E. Meidt\inst{8}
          \and
          O. V. Egorov\inst{1}
          \and
          A. T. Barnes\inst{9, 10}
          \and
          J. C. Lee\inst{11,12}
          \and
          F. Bigiel\inst{9}
          \and
          M. Boquien\inst{13}
          \and
          R. Chandar\inst{14}
          \and
          M. Chevance\inst{3,15}
          \and
          D.~A.~Dale\inst{16}
          \and
          K. Grasha\inst{17}
          \and
          R.~S.~Klessen\inst{3}
          \and
          J.~M.~D.~Kruijssen\inst{18,15}
          \and
          K. L. Larson\inst{19}
          \and
          J. Li\inst{1}
          \and
          J. E. Méndez-Delgado\inst{1}
          \and
          I. Pessa\inst{20}
          \and
          T. Saito\inst{21}
          \and
          P. Sanchez-Blazquez\inst{22,23}
          \and
          S. K. Sarbadhicary\inst{5,6}
          \and
          F. Scheuermann\inst{1}
          \and
          D. A. Thilker\inst{24}
          \and
          T. G. Williams\inst{25}
          }

   \institute{Astronomisches Rechen-Institut, Zentrum f{\"u}r Astronomie der Universit{\"a}t Heidelberg, M{\"o}nchhofstra{\ss}e 12-14, 69120 Heidelberg,
Germany
\\
              \email{elizabeth.watkins@uni-heidelberg.de}
         \and
    International Centre for Radio Astronomy Research, University of Western Australia, 7 Fairway, Crawley, 6009 WA, Australia
        \and
    Universit\"{a}t Heidelberg, Zentrum f\"{u}r Astronomie, Institut f\"{u}r Theoretische Astrophysik, Albert-Ueberle-Stra{\ss}e 2, D-69120 Heidelberg, Germany
        \and
    Space Telescope Science Institute, 3700 San Martin Drive, Baltimore, MD, 21218, USA
        \and
    Department of Astronomy, The Ohio State University, 140 West 18th Avenue, Columbus, Ohio 43210, USA
        \and
    Center for Cosmology and Astroparticle Physics, 191 West Woodruff Avenue, Columbus, OH 43210, USA
        \and
    Max-Planck-Institut f\"ur Astronomie, K\"onigstuhl 17, D-69117 Heidelberg, Germany
        \and
    Sterrenkundig Observatorium, Universiteit Gent, Krijgslaan 281 S9, B-9000 Gent, Belgium
        \and
    Argelander-Institut f\"{u}r Astronomie, Universit\"{a}t Bonn, Auf dem H\"{u}gel 71, 53121, Bonn, Germany
        \and
    European Southern Observatory, Karl-Schwarzschild-Stra{\ss}e 2, 85748 Garching, Germany
        \and
    Gemini Observatory/NSF’s NOIRLab, 950 N. Cherry Avenue, Tucson, AZ, USA
        \and
    Steward Observatory, University of Arizona, 933 N Cherry Ave, Tucson, AZ 85721, USA
        \and
    Centro de Astronomía (CITEVA), Universidad de Antofagasta, Avenida Angamos 601, Antofagasta, Chile
        \and
    Ritter Astrophysical Research Center, The University of Toledo, Toledo, OH 43606, USA
        \and
    Cosmic Origins Of Life (COOL) Research DAO, coolresearch.io
        \and
    Department of Physics \& Astronomy, University of Wyoming, Laramie, WY 82071, USA
        \and
    Research School of Astronomy and Astrophysics, Australian National University, Canberra, ACT 2611, Australia
        \and
    Technical University of Munich, School of Engineering and Design, Department of Aerospace and Geodesy, Chair of Remote Sensing Technology, Arcisstr. 21, 80333 Munich, Germany
        \and
    AURA for the European Space Agency (ESA), Space Telescope Science Institute, 3700 San Martin Drive, Baltimore, MD 21218, USA
        \and
    Leibniz-Institut f\"ur Astrophysik Potsdam (AIP), An der Sternwarte 16, 14482 Potsdam, Germany
        \and
    National Astronomical Observatory of Japan, 2-21-1 Osawa, Mitaka, Tokyo, 181-8588, Japan
        \and
    Departamento de Física de la Tierra y Astrofísica, Facultad de CC Fisicas, Universidad Complutense de Madrid, 28040, Madrid, Spain
        \and
    Instituto de Física de Partículas y del Cosmos, IPARCOS-UCM, Fac. CC. Físicas, Universidad Complutense de Madrid, Madrid, 28040, Spain
        \and
    Department of Physics and Astronomy, The Johns Hopkins University, 3400 North Charles Street, Baltimore, MD 21218, USA
        \and
    Sub-department of Astrophysics, Department of Physics, University of Oxford, Keble Road, Oxford OX1 3RH, UK
             }

   \date{Received 3 February 2023; accepted 17 May 2023}

  \abstract
   {Star formation and stellar feedback are interlinked processes that redistribute energy, turbulence, and material throughout galaxies. Because young and massive stars form in spatially clustered environments, they create pockets of expanding gas termed superbubbles, which retain information about the physical processes that drive them. As these processes play a critical role in shaping galaxy discs and regulating the baryon cycle, measuring the properties of superbubbles provides important input for galaxy evolution models.
   }
   {With the wide coverage and high angular resolution ($\sim$50--150~pc) of the PHANGS--ALMA \CO\ ({\it J}=2$-$1) survey, we can now resolve, identify and characterise a statistically representative number of superbubbles using molecular gas in nearby galaxies.}
   {We identify superbubbles by requiring spatial correspondence between shells in CO with stellar populations identified in PHANGS--HST. Then, by combining the properties of the stellar populations with the CO, we quantify the energetics of the stars and constrain feedback models. We visually find 325 cavities across 18 PHANGS--ALMA galaxies, 88 of which have clear superbubble signatures (unbroken shells, central clusters, kinematic signatures of expansion). We measure their radii and expansion velocities using CO (2--1) to dynamically derive their ages and the mechanical power driving the bubbles, which we use to compute the expected properties of the parent stellar populations driving the bubbles.}
   {We find consistency between the predicted and derived stellar ages and masses of the stellar populations if we use a supernova (SN) model that injects energy with a coupling efficiency of $\sim10$\%. Not only does this confirm that molecular gas accurately traces superbubble properties, but it also provides key observational constraints for superbubble models. We also find evidence that the bubbles are sweeping up gas as they expand, and speculate that these sites have the potential to host new generations of stars.}
  {This work demonstrates that molecular superbubbles provide novel quantitative constraints on SNe feedback efficiencies and gas clearing times, and represent a promising environment to search for the propagation of star formation, all of which are needed to understand what sets the observed star formation rates in galaxies.}

\maketitle

\section{Introduction} \label{sec:into}
Stellar feedback is the process by which the formation and evolution of high-mass stars impacts future star formation \citep{colling_impact_2018,keller_empirically_2022}. On cloud scales (5--100~pc; \citealt{miville-deschenes_physical_2017}), feedback interrupts star formation by destroying and removing the natal cold molecular gas \citep{kruijssen_fast_2019,chevance_pre-supernova_2022}. On galactic scales, it replenishes the ionised gas needed to maintain the warm phases of the interstellar medium (ISM) while injecting and redistributing material throughout galaxies via outflows and turbulence \citep{mckee_theory_1977,collacchioni_semi-analytic_2018,kreckel_measuring_2020}. Therefore, feedback plays a critical role in driving the chemical and physical evolution of galaxies and is necessary to understand why star formation is inefficient compared to the depletion time \citep{ostriker_regulation_2010,hopkins_galaxies_2014}. Because of this, accurate feedback prescriptions are vital for mapping the star formation cycle within observations and when realistically simulating this cycle across a range of physical scales (from individual star-forming cores to large-scale cosmological simulations; \citealt{tanaka_impact_2017,li_star_2018,rosen_role_2020,fensch_role_2021,grudic_dynamics_2022}). Even so, there are still many open questions about the exact time and size scales over which different feedback mechanisms dominate since observations of young star-forming regions, where stellar feedback impacts the energetics, provide only a snapshot view of these processes \citep{lopez_role_2014,mcleod_feedback_2019,barnes_which_2020-1,barnes_comparing_2021,olivier_evolution_2021}.

In addition, star formation is spatially clustered \citep{lada_embedded_2003,motte_high-mass_2018,krumholz_star_2019}. Within clustered environments, multiple feedback mechanisms occur simultaneously, which modifies how the different feedback mechanisms interact with the surrounding gas, making it more difficult to disentangle their relative importance for regulating star formation across different scales. Recent cloud-scale studies show that most gas conditions lead to quick gas dispersal before the first supernovae (SNe) occur, though if gas densities are high ($>$1\e$^4$), it can limit the effectiveness of stellar feedback until later times \citep{grasha_connecting_2018,hannon_h_2019,haid_silcc-zoom_2019,kruijssen_fast_2019,watkins_feedback_2019-1,grasha_spatial_2019,chevance_pre-supernova_2022,kim_environmental_2022}. 
Clustered star formation can even lead to the formation of hot overpressurised regions of expanding ionised gas, termed superbubbles \citep{castor_interstellar_1975,weaver_interstellar_1977,mac_low_superbubbles_1988,ostriker_astrophysical_1988}. More specifically, the feedback from multiple high-mass stars sweeps up the ISM, leaving behind hot cavities enclosed by a thin cooler shell at the boundaries. Superbubbles span a large range of spatial and temporal scales (they can reach sizes of $\sim 3$~kpc over tens of millions of years; \citealt{egorov_complexes_2017}), and therefore superbubbles, and the feedback driving them, play a significant role in shaping the ISM, which makes them great test-beds to study the impact of different feedback mechanisms on their surroundings. However, as most feedback mechanisms are time dependent, studying a single superbubble in isolation provides a limited and potentially misleading picture of how different feedback mechanisms interact. For this purpose, this work focuses on studying a sample of superbubbles detected in cold molecular gas in nearby spiral galaxies to explore the outcome of stellar feedback on physical scales of $\sim$100~pc and timescales of 0--10~Myr.

Superbubbles contain distinct features over a range of wavelengths and timescales, making them an accessible target in large surveys with different tracers \citep{churchwell_bubbling_2006,chu_x-rays_1995,bagetakos_fine-scale_2011,krause_26al_2015-1,egorov_star_2018,jayasinghe_milky_2019,pokhrel_catalog_2020,watkins_phangsjwst_2023}. Their defining feature, their shells, appear as a ring $>30$~pc in size in neutral and molecular gas tracers due to the absence of cold gas inside the superbubble and increased column densities at the bubble edge. 

The gas phases (i.e. molecular, neutral, and ionised) present when detecting the superbubble shell depend on the age of the superbubble and the surrounding gas density. At early times we expect to detect all three gas phases in the shell. As the superbubble ages, the molecular gas is quickly destroyed on timescales of 1--5~Myr, \citep{hollyhead_studying_2015,corbelli_molecules_2017,hannon_h_2019,kruijssen_fast_2019,chevance_lifecycle_2020,chevance_pre-supernova_2022,kim_star_2021,kim_environmental_2022} leaving only ionised and neutral gas behind. After $\sim$20--40~Myr the most massive stars have undergone SN explosions, removing the dominant source of ionising photons, and therefore the oldest superbubbles are usually detected using neutral gas tracers, though sometimes ionised gas can still be created at the inner edge of the shell \citep{bagetakos_fine-scale_2011,egorov_complexes_2017}. At earlier stages ($\sim$10~Myr) the cavity itself is filled with ionised gas arising from photoionisation and shocks from stellar winds and SNe. In addition, the stellar winds that shock heat the gas produce X-ray emission, which can be detected in combination with the bubble shell morphology \citep{lopez_what_2011}. 

The final distinguishing feature of superbubbles are their expansion motions, typically reaching a few tens of \kms, depending on the age, ambient density, and source of energy. For superbubbles that have not stalled, as the shells expand a detectable kinematic feature is created in emission line tracers, which can also be used to identify superbubbles, estimate their ages, and potentially determine the source of the feedback driving the expansion. In this work, it is the detection of expansion motions that ultimately allows us to distinguish superbubbles from \HII\ regions, which have much lower expansion velocities ($\sim$2.5~\kms; \citealt{tremblin_age_2014}) due to the vastly different amounts of energy injected into the gas by the feedback mechanisms driving \HII\ regions versus\ superbubbles (photoevaporation vs\ winds and SNe, respectively).\footnote{If stalled, superbubbles usually do not have detectable expansion motions. However, they are usually distinct from \HII\ regions morphologically due to their larger size ($>30$~pc) and lack of concentrated ionised gas emission within their cavity.}

In nearby galaxies superbubbles are normally surveyed using 21~cm line emission from \HI\ since \HI\ is present for almost the entirety of the bubbles' life (except the earliest stages before its molecular gas has been dissociated; \citealt{oey_h_2002,bagetakos_fine-scale_2011}), and is present in the outer discs of galaxies. \HI\ line emission also provides kinematic constraints on bubble properties, particularly on large (kiloparsec) scales tracing the morphology and shape of older ($>$20~Myr) bubbles \citep{bagetakos_fine-scale_2011}. However, \HI\ observations are less sensitive for the inner parts of massive spiral galaxies since most of the gas is molecular in these environments (though this is not true in dwarf galaxies). Therefore, superbubbles detected with \HI\ in such galaxies trace the impact of feedback on gas that is not actively forming stars (i.e. they trace timescales longer than the star-forming timescale). In addition, it can sometimes be unclear if an \HI-traced superbubble is feedback-driven or is instead a dynamically created hole. With old superbubbles, it is difficult to identify the driving stellar populations powering them since the clusters dynamically decouple from the gas, causing them to move away, and the brightest stars (i.e. the O-stars) are the first to undergo SNe \citep{warren_formation_2011}.

Superbubbles can also be traced using ionised gas tracers, such as \Ha\ \citep{sanchez-cruces_kinematics_2015-2,camps-farina_physical_2017,gerasimov_stellar_2022}. This comes with the advantage of being able to directly measure feedback pressures using the ionised gas properties, providing a direct estimate of the dominant feedback mechanisms. However, it does not trace the cold dense molecular gas that stars form from, and so we are limited to tracing the strength of feedback rather than the direct impact of feedback on the ISM's ability to form stars. Furthermore, the velocity resolution of optical line emission is often too coarse to directly observe expansion of superbubbles, let alone the expansion of \HII\ regions. Though we note here that the expansion velocity can be recovered using velocity dispersion measurements \citep{smirnov-pinchukov_measurements_2021}, which can then be used to differentiate between ionised emission from \HII\ regions and superbubbles that have not yet stalled (Egorov et al., subm.).

The final gas phase left to trace superbubbles is molecular gas. If the surrounding gas is sufficiently dense ($\gtrsim50$~cm$^{-3}$), or if the superbubble formed recently enough that the gas has not been dissociated by ionising radiation, the swept-up shell will still contain molecular gas. However, studies that investigate molecular bubbles are usually limited to small, nearby bubbles in the Milky Way (up to a few tens of pc) or instead focus on \HII\ regions \citep{arce_bubbling_2011}. There are three reasons for this. First, line-of-sight effects in the Milky Way make detecting large molecular superbubbles difficult, and limit molecular superbubble studies to the inner part of the galaxy ($<$8~kpc) around the Sun. More often, they are limited to the local neighbourhood ($<$1~kpc) around the Sun \citep{ochsendorf_nested_2015,joubaud_gas_2019,zucker_star_2022}, whereas \HI\ and ionised gas tracers can be used to detect superbubbles over a larger volume of the Milky Way (\citealt{ehlerova_h_2005} \citealt{ehlerova_hi_2013}). Second, molecular gas has a small filling factor. Finally, molecular gas is quickly destroyed when exposed to ionising radiation within 1--5~Myr, especially when densities are insufficient to shield the gas \citep{hollyhead_studying_2015,corbelli_molecules_2017,hannon_h_2019,kruijssen_fast_2019,chevance_lifecycle_2020,chevance_pre-supernova_2022,kim_star_2021,kim_environmental_2022}.
Therefore, the time frame for detecting molecular gas around bubbles is short, limiting the maximum time and size scales we can detect them compared to H\textsc{i} (a few 100 pc and up to timescale of $\sim$10~Myr \citealt{nath_size_2020}). The fact that lower-resolution molecular gas studies of nearby galaxies rarely detect superbubbles (and when they do, they are $>700$~pc \citealt{tsai_molecular_2009-1}) highlights the need for observations at 100~pc resolution. 

However, the shorter timescales are advantageous for studying the impact of feedback directly on the surrounding ISM in nearby galaxies, ensuring that we measure properties from a younger population of superbubbles if we reach the physical resolution needed to detect them since younger superbubbles should be smaller unless they are powered by a large stellar
population ($>10^5$~\msun).
By tracing superbubbles at earlier stages, we can directly measure the impact that expanding superbubbles have on gas that is actively forming stars, especially at $<0.5r_{25}$ ($r_{25}$ is defined as the \textit{B}-band isophote at 25~mag~arcsec$^{-2}$) where there is less \HI\ because most of the gas is molecular. Moreover, the molecular gas around superbubbles might contain material that forms the next generations of stars via sequential star formation processes \citep{elmegreen_sequential_1977}. By focusing on nearby galaxies, we also retain the galactic context in which molecular superbubbles form, allowing us to link the properties of the local gas into which the bubbles expand to the larger-scale galactic environment. However, the restrictions imposed by quick molecular gas destruction mean high angular resolution observations (at least 100~pc) over large areas are needed to resolve and detect a significant number of superbubbles. Consequentially, no dedicated surveys exist investigating molecular superbubbles in nearby galaxies, and currently there are only a small number of studies where molecular superbubbles were detected, all of which use interferometry and focus on central star-bursting regions \citep{sakamoto_molecular_2006, tsai_molecular_2009-1, bolatto_suppression_2013}. 

With the onset of ALMA we can now, for the first time, map molecular gas over large areas of nearby galaxies at high spatial and spectral resolution. We note here that while higher-resolution observations are available with JWST with the PHANGS\footnote{Physics at High Angular resolution in Nearby GalaxieS} \url{http://phangs.org}--JWST Treasury program \citep{lee_phangsjwst_2023}, only part of the survey has been completed, and so we defer adding JWST observations until after the PHANGS--JWST survey is complete. Therefore, we use the PHANGS--ALMA CO (J=2$-$1) maps of nearby galaxies with coincident PHANGS--MUSE and PHANGS--HST data (see Section~\ref{sec:obs}) to catalogue a sample of molecular superbubbles in Section~\ref{sec:method} large enough to provide baseline expectations for identifying and analysing molecular superbubbles in nearby ($<20$~Mpc) galaxies. In Section~\ref{sec:general-results} we measure the basic properties of the bubbles (radius, mass) using CO, and leverage HST observation to identify the true stellar populations driving the bubbles. Combining CO properties with the stellar populations allows us to constrain feedback models and their efficiencies in Section~\ref{sec:derived-props}. In Section~\ref{sec:discuss} we discuss the origin of the bubbles and the mass within them, and what factors lead to detecting them. Finally, in Section \ref{sec:conclude} we summarise our findings. 

\begin{table}[]
    \caption{List of the PHANGS--ALMA galaxies studied in this work, their properties, and the number of superbubbles found within them from \cite{leroy_phangsalma_2021}. SFR is the star formation rate, $M_\text{star}$ is the stellar mass, $M_{\text{H}\textsc{i}}$ is the mass of neutral hydrogen, and $M_{\text{H}_2}$ is the mass of molecular hydrogen for the entire galaxy.}
\begin{threeparttable}
\resizebox{\columnwidth}{!}{

\begin{tabular}{rrrrrrr}
Galaxy   & Distance\tnote{a} & No. of     & SFR               & $M_\text{star}$ & $M_{\text{H}\textsc{i}}$ & $M_{\text{H}_2}$  \\
         & (Mpc)    & bubbles   & (\msun~yr$^{-1}$) & ($10^{9}$\msun) & ($10^{9}$\msun)          & ($10^{9}$\msun)  \\
\hline
NGC~0628 & 9.84     & 34 & 1.8 & 21.9 & 5.0  & 4.7  \\
NGC~1087 & 15.85    & 15 & 1.3 & 8.6  & 1.3  & 1.7  \\ 
NGC~1300 & 18.99    & 11 & 1.2 & 41.4 & 2.4  & 3.2  \\
NGC~1365 & 19.57    & 6 & 16.9 & 97.8 & 8.7  & 24.6 \\
NGC~1385 & 17.22    & 12 & 2.1 & 9.5  & 1.6  & 1.8  \\
NGC~1433 & 18.63    & 12 & 1.1 & 73.4 & 2.5  & 2.7  \\
NGC~1512 & 18.83    & 15 & 1.3 & 51.6 & 7.6  & 1.9  \\
NGC~1566 & 17.69    & 31 & 4.5 & 60.9 & 6.4  & 6.2  \\
NGC~1672 & 19.40    & 13 & 7.6 & 53.6 & 16.0 & 9.1 \\
NGC~2835 & 12.22    & 6  & 1.2 & 10.0 & 3.0  & 1.0 \\
NGC~3351 & 9.96     & 13 & 1.3 & 23.0 & 0.8  & 1.9 \\
NGC~3627 & 11.32    & 39 & 3.8 & 68.1 & 1.2  & 7.0 \\
NGC~4254 & 13.10    & 37 & 3.1 & 26.6 & 3.0  & 8.2 \\
NGC~4303 & 16.99    & 24 & 5.3 & 33.4 & 4.6  & 11.4 \\
NGC~4321 & 15.21    & 25 & 3.6 & 55.6 & 2.7  & 9.7 \\
NGC~4535 & 15.77    & 16 & 2.2 & 34.0 & 3.7  & 7.1 \\
NGC~5068 & 5.20     & 6  & 0.3 & 2.5  & 0.7  & 0.4 \\
NGC~7496 & 18.72    & 10 & 2.3 & 9.9  & 1.2  & 2.1 \\

\end{tabular}
}
\begin{tablenotes}\footnotesize
\item[a] Distances from \citet{anand_distances_2021}, also see acknowledgements.

\end{tablenotes}
\end{threeparttable}
    \label{tab:sum}
\end{table}

\section{Observations} \label{sec:obs}
In this section we provide a brief overview of the PHANGS--ALMA, PHANGS--MUSE, and PHANGS--HST data sets and associated products used in this study. We used 18 galaxies from the sample (details listed in Table \ref{tab:sum}) that were observed by all three of the PHANGS large programs (with detectable \CO) for superbubble signatures. We used ALMA \CO\ data cubes and associated moment maps and peak intensity temperature maps, the MUSE optical emission line maps \citep{emsellem_phangs-muse_2022}, and the HST \textit{B}-band data sets \citep{lee_phangs-hst_2022} and multi-scale stellar association catalogues \citep{larson_multi-scale_2022}.

\subsection{PHANGS--ALMA} \label{sec:obs-alma}
\noindent The PHANGS--ALMA survey consists of 90 galaxies mapped with \CO\ (J=2$-$1) (hereafter CO) using the 12m, 7m, and Total Power (TP) ALMA arrays with an angular resolution of $\sim\ang{;;1}$, reaching physical scales of $\sim$50--150~pc throughout the sample, and a velocity resolution of 2.5~\kms. We use the combined 12m+7m+TP maps (public release v1.0), which include single-dish data and so are sensitive to all spatial scales. This provides us with the high resolution needed to resolve and characterise superbubble properties with CO while preserving the extended emission needed to correctly measure the total emission present within the superbubbles. The exact details of the reduction pipeline, references, and survey strategy are provided in \cite{leroy_phangsalma_2021-1,leroy_phangsalma_2021}, but we provide a brief summary of the reduction steps and subsequent product production here. 

The combined 12m+7m+TP maps were made using the PHANGS--ALMA pipeline. The PHANGS--ALMA pipeline first imaged the 12m+7m data sets together using a multi-scale clean cycle followed by a single-scale clean cycle, 
while the TP observations were imaged separately. After the combined 12m+7m maps were primary beam corrected and convolved to produce a round Gaussian-shaped beam, the 12m+7m and TP were feathered together in Fourier space. If any galaxies were observed with multiple mosaics (required for galaxies with $>$150 pointings), they were reduced separately and linearly combined into a single mosaic at the end. Along with the native resolution data cubes, each cube was convolved at seven fixed physical spatial scales (between 60 to 1000~pc) to produce observations at matching scales.

From these cubes, the pipeline constructs a 3D noise model and then propagates this error into the subsequent moment maps (moment 0--2 maps and additional analysis maps such as peak brightness temperature maps, termed $T_\text{peak}$). The moment maps were made using two masking schemes, one that prioritises high completeness, termed `broad' maps, and the other that prioritises low false-positive rates, termed `strict' maps. The strict scheme masked when $\geq$2 consecutive channels had a S/N >4. It also masked enclosed regions where $\geq$2 channels had a S/N >2 under the condition they contained at least one pixel masked at the former strict level. The broad masking scheme masks the union of all strict masks generated for all of the spatially convolved cubes (i.e. 60--1000~pc). Since we wanted to detect the faint diffuse emission, we opted to use the native (rounded-beam) CO cubes and noise maps and the broad moment maps.

\subsection{PHANGS--MUSE} \label{sec:obs-muse} 
The PHANGS--MUSE sample contains 19 galaxies imaged over 168 pointings using the Multi Unit Spectroscopic Explorer (MUSE \citealt{bacon_muse_2010-1}) instrument at the Very Large Telescope (VLT). Observations were taken using the 
wide field mode over the nominal wavelength range at a resolution in the range $\ang{;;0.56}$--$\ang{;;1.25}$ and a mean spectral resolution of $R=3000$. The exact details of their reduction and data product generation are presented in \cite{emsellem_phangs-muse_2022}, but we provide a quick overview here. Almost all pointings were observed four times each with a 90$^{\circ}$ rotational offset to minimise instrumental artefacts, especially in the velocity maps. The pointings for each galaxy were reduced using \texttt{pymusepipe} \citep{emsellem_phangs-muse_2022}, a dedicated Python wrapper that runs the MUSE data reduction recipes, MUSE--DRS \citep{weilbacher_data_2020} via the \texttt{esorex} framework, in an almost automatic fashion. The pipeline organises the observation set into science and calibration categories; runs the MUSE-DRS recipes for artefact removal, calibration, and sky subtraction; and provides functionality for semi-automatic pointing alignment and mosaicking. The final outputs are native resolution data cubes where each MUSE pointing has a unique point spread function (PSF).

The data were processed using a dedicated PHANGS--MUSE data analysis pipeline (DAP) to compute a series of products (e.g. continuum-subtracted line maps, stellar kinematics, and populations). Gaussian line profiles are fit to kinematically tied emission lines, including the bright \Ha\ line, and are corrected for foreground Milky Way extinction. This work uses the \Ha\ integrated emission line map calculated using the MUSE data analysis pipeline to help locate and identify superbubbles. We use the native resolution maps for this work since we do not need matched resolution images. We also note that we do not use IC~5332 within this work since there was not enough CO emission detected within IC~5332 to identify superbubbles.

\begin{figure*}
\centering
\includegraphics[width = \textwidth]{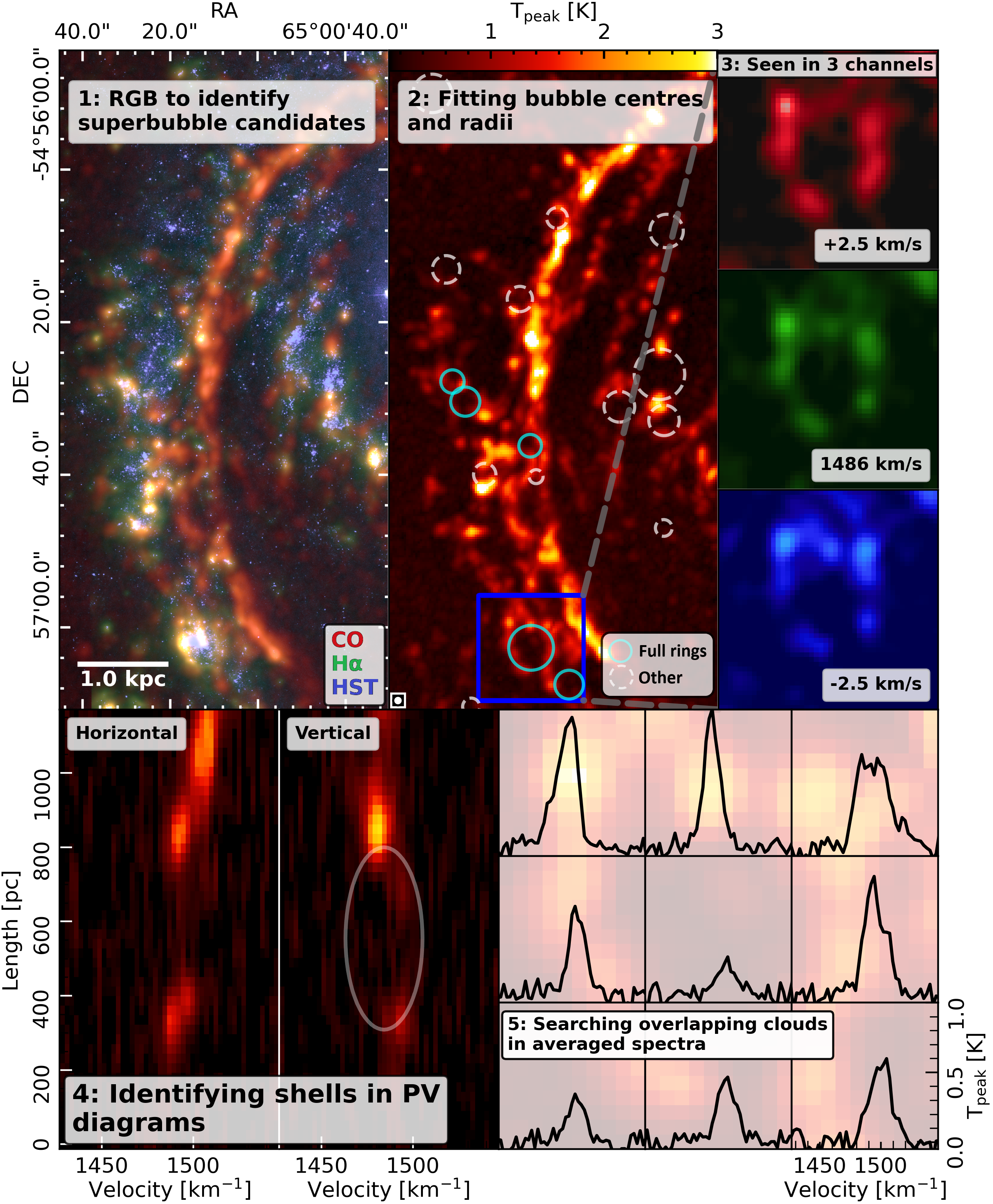}
\caption{Illustrating bubble identification and elimination criteria in section of NGC~1566. \textbf{1.} CO \Tpeak\ (red), \Ha\ (green), and HST \textit{B}-band (blue) combined into a false-colour image at their original resolution to identify superbubbles using multi-wavelength information. \textbf{2.} Manually fitting radii and their centres using the CO \Tpeak\ map. The cyan and dashed white circles show catalogued bubbles that were analysed or ignored, respectively. The blue box outlines the bubble examined in all remaining panels (Bubble 36 in Table~\ref{tab:props}). \textbf{3.} Investigating the emission across three neighbouring channels in CO to confirm if bubble emission is significant in multiple consecutive velocity bands. If not, the bubble is removed from the sample. \textbf{4.} Horizontal and vertical PV diagrams to confirm that expansion signature is present. The grey ellipse shows the present expansion signature. If unconfirmed, the bubble is removed. \textbf{5.} Illustrating average spectra around the bubble. If identifiable background or foreground emission is found, the bubble is removed. All spectra shown here are free of contaminating emission.}
\label{fig:method}
\end{figure*}

\subsection{PHANGS--HST} \label{sec:obs-hst}
PHANGS--HST observations comprise five bands of optical imaging (\textit{NUV}: WFC3/F275W, \textit{U}: WFC3/F336W, \textit{B}: WFC3/F438W, \textit{V}: WFC3/F555W, \textit{I}: WFC3/F814W) of 38 spiral galaxies at a resolution of $\ang{;;0.08}$. The galaxies previously observed with ACS bands (\textit{B}: ACS/F435W, \textit{V}: ACS/F555W, \textit{I}: ACS/F814W) as part of the LEGUS survey \citep{calzetti_legacy_2015} were not re-observed, but they were included and reduced using the PHANGS--HST data pipeline \citep{lee_phangs-hst_2022}. All galaxies used in this work had additional source catalogue products generated. Below we provide a short overview of the data reduction and catalogue generation process (for more details on these methods, see \cite{thilker_phangs-hst_2021}, \cite{turner_phangs-hst_2021} \cite{whitmore_star_2021}, \cite{lee_phangs-hst_2022}, \cite{deger_bright_2022}, and \cite{larson_multi-scale_2022} and references therein).

The HST imaging data were first drizzled and mosaicked using the current standard procedures. All bands were astrometrically aligned to the ICRS frame and a common pixel grid using the \textit{V}-band 555~nm image as a reference. To generate catalogues, as a starting point, the \textit{NUV-U-B-V-I} images were processed using \texttt{DOLPHOT} \citep{dolphin_dolphot_2016} to identify point-like sources brighter than >3.5$\sigma$. To capture sources more extended than \texttt{DOLPHOT} can reliably identify, \texttt{DAOStarFinder} \citep{stetson_daophot_1987,bradley_astropyphotutils_2022} was also run with a slightly broader 2.5 pixel FWHM kernel and selects sources with a S/N > 4. These source lists were used to generate two catalogues: compact clusters \citep{thilker_phangs-hst_2021, whitmore_star_2021} and multi-scale stellar associations \citep{larson_multi-scale_2022}. \texttt{DOLPHOT} and \texttt{DAOStarFinder} sources were used when identifying compact clusters, whereas only \texttt{DOLPHOT} sources were used to generate stellar association catalogues.

Compact cluster catalogues were generated via candidate selection based on the multiple concentration index (MCI) \citep{thilker_phangs-hst_2021} to exclude \texttt{DOLPHOT} and \texttt{DAOStarFinder} sources inconsistent with expectations for clusters at the distance of each galaxy, followed by morphological classification of cluster candidates \citep{whitmore_star_2021}. Classification was conducted via human inspection for the 1000 brightest candidates per galaxy, and subsequently a classification based on machine learning (ML) was applied to the entire set of (fainter) candidates. Accepted clusters were assigned to three categories based on their shape: 1) single-peaked symmetric clusters; 2) single-peaked asymmetric clusters; 3) multi-peaked asymmetric clusters. \cite{whitmore_star_2021} showed that human and ML classification methods produce similar catalogues. These clusters represent the smallest, brightest, and densest hierarchical structures within the galaxy, and thus exclude extended stellar mass. 

The multi-scale stellar association catalogues (\citealt{lee_phangs-hst_2022} and \citealt{larson_multi-scale_2022}) instead include the spatially distributed stellar population at 16, 32, and 64~pc scales, chosen to match the spatial scales traced with PHANGS--ALMA. They are generated using a watershed segmentation algorithm on tracer maps of the \texttt{DOLPHOT} sources identified in either the \textit{NUV}- or \textit{V}-band and smoothed to the above-mentioned spatial scales with Gaussian kernels. Stars belonging to a common overdensity (clump) found to be over a given level after the smoothing are merged into a single catalogue object (i.e. a stellar association). As the spatial scale is increased, a greater number of small-scale stellar groupings are eventually merged, with the union of associations defined at 16, 32, and 64 pc scales representing a hierarchy of objects.

Spectral energy distribution (SED) fitting with \texttt{CIGALE} \citep{boquien_cigale_2019} was used on both catalogues to calculate their masses, ages, reddenings, and the associated uncertainties. The fitting was performed using a set of single stellar populations models of different ages from \cite{bruzual_stellar_2003} at solar metallicity using a Chabrier IMF \citep{chabrier_galactic_2003}, allowing for internal extinction using \cite{cardelli_relationship_1989}. Since \texttt{CIGALE} was run using a grid, we note here that all ages are given to the nearest integer Myr (for further details, see \citealt{lee_phangs-hst_2022} and \citealt{turner_phangs-hst_2021}). 

We used the \textit{B}-band (435/438~nm) to help identify superbubble candidates and the \textit{V}-band 16--64~pc multi-scale stellar associations (internal release v1.3) to provide a benchmark for cluster properties derived using CO observations.

\section{Identifying superbubbles} \label{sec:method}
\subsection{Superbubble definition}

Bubbles are generally identified manually using a single tracer sensitive to their shell morphology \citep{bagetakos_fine-scale_2011}, although when available, two tracers are used, one sensitive to the shell and the other sensitive to emission from ionised gas that is contained within the bubble \citep{jayasinghe_milky_2019,watkins_phangsjwst_2023}. Together they help physically determine whether the superbubble is real, that is, driven by stellar feedback, rather than a hole created by turbulence, dynamic effects, or chance alignment of other structures. If velocity information is available, bubbles can also be confirmed by detecting their expansion signatures. However, it is rare to see all of these characteristic signatures in an idealised way. For instance, the bubble shells can appear in drastically different forms. They can be whole, broken, elliptical, or asymmetric; can exist as part of a larger bubble complex; or can be found embedded within larger-scale emission.

For this study we measured the properties of molecular superbubbles with HST sources with known masses and ages. Our goals were twofold: \textbf{(a)} to confirm that a significant sample of molecular superbubbles can be identified at $\sim$100~pc resolution and \textbf{(b)} to determine what drives them and how efficiently energy is injected into their shells. Both points remain a source of uncertainty for theoretical models. To achieve these goals we needed a reliable sample of molecular superbubbles. Consequentially, visual searches were preferred over automated methods since they are robust against complex bubble structures, while simultaneously weighing in on the co-spatial multi-wavelength information. For these reasons, we opted for a manual approach to identifying superbubbles. 

To maximise identifying real superbubbles, we limited the galaxy sample to those with ALMA data that we used to identify bubble shells, HST to reinforce that the bubble is driven by stellar sources, and MUSE to trace ionised gas inside bubbles. This limits the search to 18 galaxies. For these data sets, we also had to choose the specific maps. We chose ALMA CO peak temperature maps, MUSE H$\alpha$ maps, and HST \textit{B}-band maps (see Fig.~\ref{fig:method}, panel 1).
During initial testing, ALMA CO peak temperature (see Fig.~\ref{fig:method}, panel 2) had a higher contrast compared to the moment-0 CO maps, making the shell-like morphology stand out against the background.

\begin{figure*}
\centering
\includegraphics[width = \textwidth]{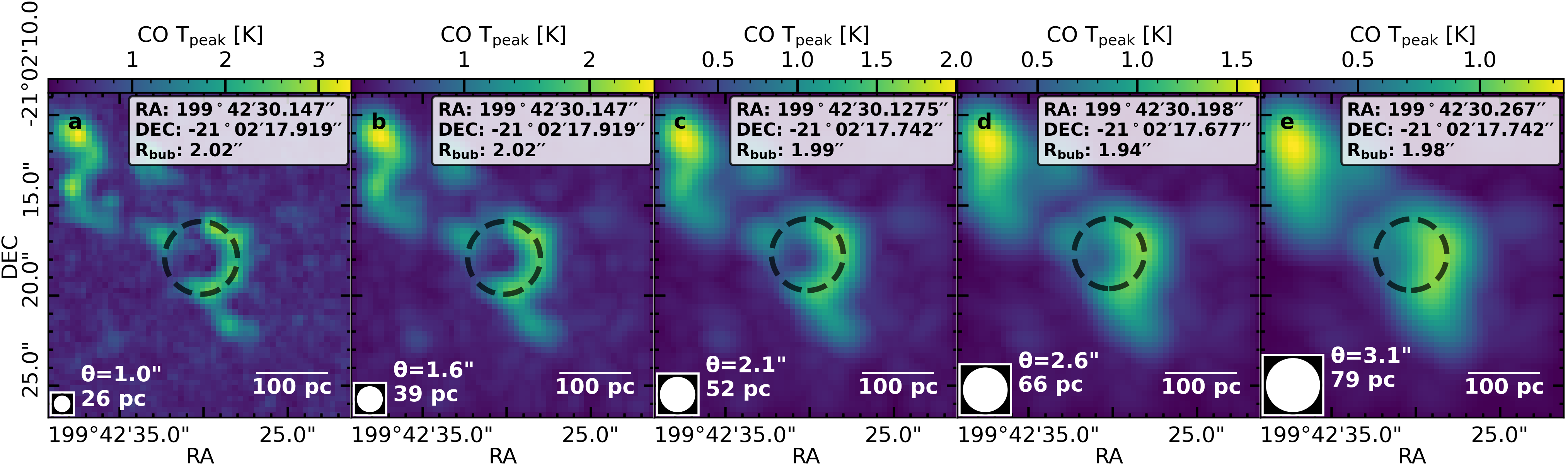}
    \caption{Bubble 82 in NGC~5068 convolved to lower spatial resolutions from left to right. Panel {\bf a} shows the superbubble at the original resolution. Each physical resolution has been refit with a new bubble radius (dashed black circle). Values of refitted radius (including the bubble centre) shown on each subplot. The filled white circle at the lower left of each panel shows the beam. The text immediately to the right of the beam shows the physical and angular resolution of the beam. The white horizontal line illustrates physical scales of 100~pc.}
\label{fig:convolve}
\end{figure*}

The \textit{B}-band HST was chosen to trace the young stellar population. While the \textit{NUV}-band traces even younger populations, the observations also suffer more from internal extinction; considering that we are searching for bubbles with significant CO where we expect higher extinction values, this might be a problem and cause us to miss sources. MUSE H$\alpha$ is the strongest optical emission line tracing warm ionised gas, making it the obvious choice. However, assuming the bubbles formed from a single burst of star formation, MUSE H$\alpha$ limits the age range of bubbles we detect to $<$10~Myr, which is the expected emission lifetime of \Ha\ \citep{whitmore_using_2011,haydon_uncertainty_2020}.\footnote{This is a soft limit since we actually expect stars to form over an extended period of time rather than in an instantaneous burst, making superbubbles multi-generational objects.} However, for our study, it is advantageous as it provides strong evidence linking the recent star formation to the bubble structure and HST stellar population. More importantly, these timescales are also consistent (and usually longer) with recent estimates for the lifetime of molecular gas after the onset of star formation (i.e. 1--5~Myr \citealt{chevance_lifecycle_2020,chevance_pre-supernova_2022,kim_environmental_2022}). As a result, we expect CO will ultimately limit our ability to detect bubbles, rather than missing \Ha\ emission.

\subsection{Identification method}
The initial search and quality assurance was undertaken by EJW using tiles of roughly $\ang{;;60}\times\ang{;;60}$ (although this was not a strict rule), and performed twice per galaxy. All three images (ALMA, MUSE, HST) were stacked for each galaxy as an RGB image, shown in panel 1 of Fig.~\ref{fig:method}. We used a square root stretch for the CO to help emphasise weaker emission features, a log stretch for \Ha\ since the emission spans many orders of magnitude, and a linear stretch for HST to focus on the bright point-like sources. We performed an exhaustive search for any round ring-like features in CO (including partial rings) at any size scale and thickness that had concentrated co-spatial HST sources (i.e. cluster-like sources and point sources) and concentrated \Ha\ emission at the centre or the edge of the ring (see Fig.~\ref{fig:method}, panel 1). Features matching these three criteria are initially selected. Focusing on concentrated emission typically causes us to exclude extremely large, partial CO shells that are likely older superbubbles (usually $>300$~pc and $>10$~Myr). We initially tried to include older superbubbles to trace a wider range of evolutionary stages, but we were unable to accurately determine their expansion velocities in PV space. Without a consistent estimate of their expansion velocity, we had to reject them.

After identifying a superbubble candidate, we checked their velocity structures by scanning through the CO channel map. All shells persisted in $\geq$3 channels (corresponding to $\sim$7.5~\kms), and therefore we considered them a significant detection (Fig.~\ref{fig:method}, panel 3). We next plotted vertical and horizontal position velocity (PV) diagrams (Fig.~\ref{fig:method}, panel 4) and integrated velocity spectra in a radial pattern (Fig.~\ref{fig:method}, panel 5) to perform a more detailed check of the velocity structure. In the PV diagrams, bubble expansion appears as a hole or an arc if one side has blown out or if the emission is too weak to be detected. If a bubble is expanding, integrated velocity spectra can sometimes show evidence of this as double-peaked spectra or as small wings in emission spectra \citep{camps-farina_physical_2017}, though after testing we found it was uncommon to see expansion signatures in the spectra wings, so instead we used them to identify overlapping emission. Together, these steps allowed us to confirm the presence of kinematic signatures indicative of bubble expansion and to remove bubbles significantly contaminated by overlapping velocity features, such as multiple clouds along the line of sight. If we were unable to identify the bubble in PV space, or if any were significantly contaminated by multi-component emission features, we excluded the bubble from the sample. Not only did this remove less reliable candidates, but it also removed bubbles unsuited for further analysis because without a reliable measure of expansion velocity, we could not derive the dynamical age or the mechanical energy injected into the gas.

In total, we rejected 25 bubbles based on their PV diagrams, yielding a final sample of 325 bubbles. The number of superbubbles found per galaxy is listed in Table \ref{tab:sum}. 

\subsection{Fitting bubble radii}
\begin{figure}
\centering
\includegraphics[width = \columnwidth]{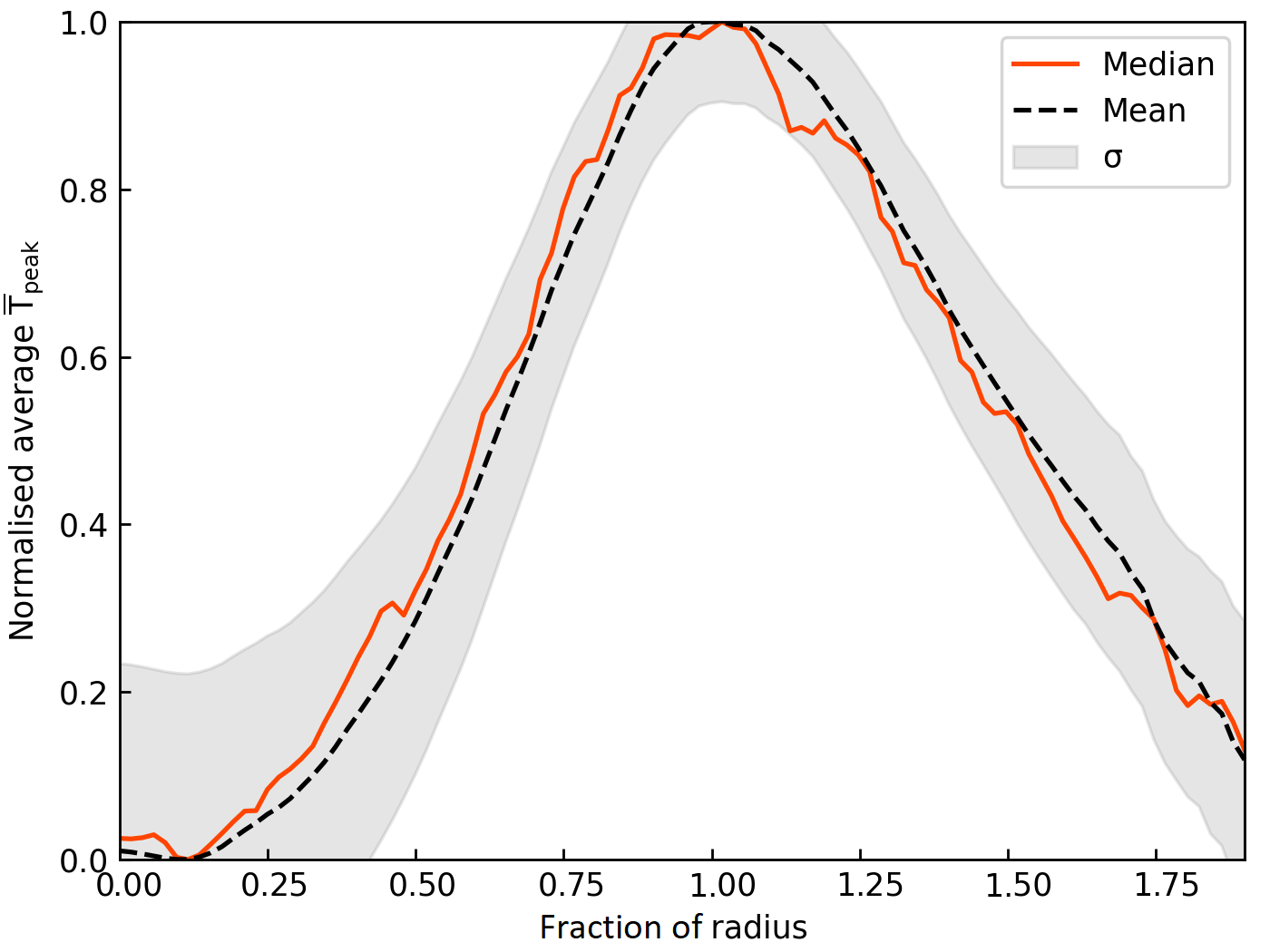}
    \caption{Averaged normalised intensity profile of bubbles categorised as  Closed Rings. The solid red line shows the mean profile, the dashed black line shows the median profile, and the filled grey region shows the statistical spread of the data measured using the standard deviation ($\sigma$).}
\label{fig:profile}
\end{figure}

The physical properties of the superbubbles (i.e. bubble centres and radii) were fit visually by drawing circular apertures. We note here that we did not exclude elliptical features, but we found that all the bubble features identified were well characterised with a simple circular aperture. To fit the circular aperture, we used the peak emission of the shells (i.e. the shell ridge) using CO peak temperature rather than the inner edge, which is what superbubble models use instead. Not only are the shell ridges easier to identify visually, but our choice minimises resolution uncertainties. The shells of idealised superbubbles are expected to be thin, due to the cooling in the propagating shock front ($\sim$0.1~pc), and if the cooling results in a thick shell, models and observations typically observe thicknesses of up to $\sim10$~pc \citep{ochsendorf_nested_2015, krause_surround_2018, joubaud_gas_2019}. We find the thickness of the shells usually matches the ALMA beam (see panel 2 of Fig.~\ref{fig:method}), and therefore they are unresolved. More importantly, the peak intensity can be consistently measured in the same location since it is always defined by the highest intensity. When unresolved, the inner edge will vary depending on the physical resolution reached, meaning no clear inner edge exists.

To test the impact resolution might have on fitting apertures to superbubbles, we took our most resolved and nearly perfect bubble candidate (Bubble 82 in NGC~5068), convolved the data cube, and remade the \Tpeak\ intensity map. We convolved the bubble up to three times its original resolution (a factor of three was the point where the bubble structure was no longer visible in CO since the beam was comparable in area to the bubble itself) in 20 steps (i.e. 1.1, 1.2, 1.3\dots3.0) and remeasured the radius, which we illustrate in Fig.~\ref{fig:convolve}. We find that the convolution did not significantly affect the measured size of the bubble, and at most slightly reduced the radius we measured. We did find, however, that the central position of the bubble was more difficult to define and more likely to be offset compared to the higher angular resolution data, due to the bottom left side of the bubble containing less emission, affecting the perceived centre of the bubble. This suggests that bubbles with resolved ring structures are not systematically larger at lower physical resolutions, though the bubble centres might be more uncertain.

To quantify our level of uncertainty in visually identifying the shell ridge of bubbles, we show the average normalised change in intensity as a function of radius in Fig.~\ref{fig:profile} for a subset of the sample containing unbroken shell morphologies ( Closed Rings, defined below). To do this, we first calculated the radially averaged $T_\text{peak}$ intensity profile normalised between 0 and 1 of each bubble after normalising their radii to 1 at $R=1$, where $R$ is the bubble radius. We then took the average of the entire sample and re-normalised the intensity to produce a single-intensity profile. Figure~\ref{fig:profile} shows that the intensity decreases for radii larger and smaller than one, and the intensity is at a minimum at the centre confirming that on average, the bubble parameters are correct. The average profile also indicates that a $\sim$10\% change in radii size yields little change in the intensity. We also note that when remeasuring the radii in Fig.~\ref{fig:convolve}, the change in measured radii from repeated measurements was always within this $\sim$10\% margin. These results indicate that our bubble sizes are accurate to within approximately 10\%.

\subsection{Quality check via peer review}
\begin{figure}
\centering
\includegraphics[width = \columnwidth]{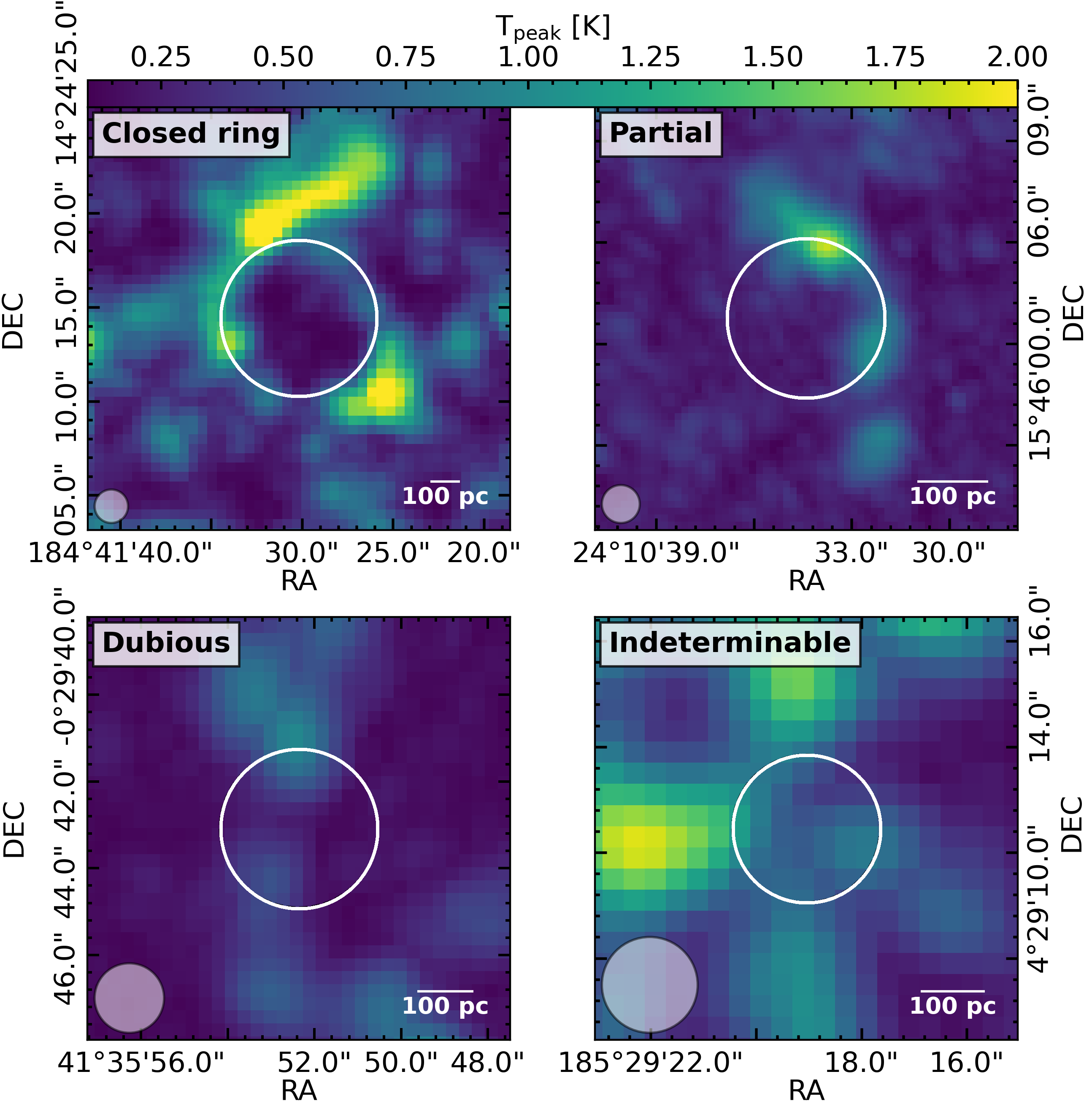}
    \caption{Four typical bubbles identified and categorised as  Closed Ring (upper left), partial superbubble (upper right), Dubious superbubble (lower left), and Indeterminable (lower right) shown in CO peak temperature with their radii shown as white apertures. The translucent grey circle at the bottom left indicates the beam, while the white horizontal line indicates the physical size.}
\label{fig:bub-type}
\end{figure}

To confirm that the superbubbles we identify are visible to others, we ranked the bubble candidates using four members of the team. Since this work presents the first significant sample and analysis of molecular superbubbles in nearby galaxies, we focus on identifying bubbles that are robust in CO only as a measure of their reliability. Therefore, we categorised each bubble as follows: \textbf{1.}  Closed Rings for near-perfect enclosed rings (cyan apertures in Fig.~\ref{fig:method}, panel 2); \textbf{2.}  Broken Rings for partial or incomplete rings; \textbf{3.} Dubious for bubbles that cannot be seen or that are unconvincing in CO; and \textbf{4.} Indeterminable for bubbles that cannot be placed in the other categories. All dashed white apertures in Fig.~\ref{fig:method}, panel 2, illustrate the last three categories, and in Fig.~\ref{fig:bub-type} we show an example of a bubble from each of the four categories. If two or more people agreed on the classification, we assigned that classification. If tied, the better of the two classifications was assigned, otherwise, we assigned Indeterminable for no agreement. In total 88 bubbles are labelled as  Closed Rings, 167 as   Broken Rings, 65 as Dubious, and 5 as Indeterminable, with only 1 bubble in this classification for which there was no agreement on which category to place the bubble in. Almost all bubbles identified as Dubious are very small, and are nearly the same size as the image resolution. These represent bubbles that are unlikely to be detected as a superbubble without overlapping HST and MUSE data. The  Broken Rings contain a much larger fraction of bubbles with an odd morphology or that have larger radii. Therefore, these likely represent molecular superbubbles where their CO has been destroyed or is in the process of being destroyed. 

\begin{figure*}
\centering
\includegraphics[width = \textwidth]{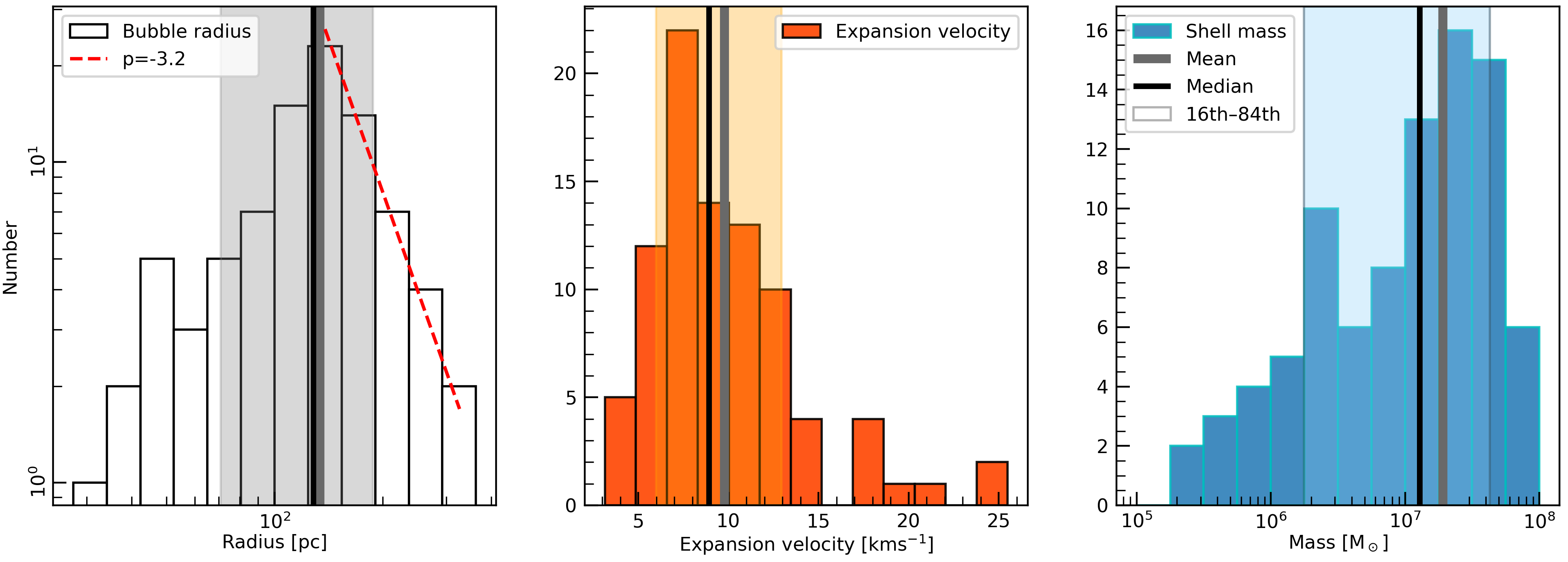}
    \caption{Histogram distributions of the bubble radii \textit{(left)}, expansion velocity \textit{(middle),} and shell mass \textit{(right)} for all 88 superbubbles. The dashed red line in the left panel shows a power-law fit to the radii with an exponent of $-3.2$. The thick vertical grey and black lines show the mean and median of each distribution, respectively, and the coloured filled regions show the 16{th}--84{th} sigma percentiles.}
\label{fig:meas-bub_props}
\end{figure*}

\subsection{Final sample}
For the final sample we focus on the  Closed Rings, which represent the cleanest superbubble examples. These allow us to reliably determine the mechanical energy injected into the gas by the central stellar cluster. Currently, this efficiency is not well known since it is hard to measure observationally. We excluded the Dubious and Indeterminable morphologies since we could not confirm if they were real visually using only CO. We also excluded  Broken Ring morphologies. While the  Broken Rings are real, we cannot determine if their broken morphology is a result of them expanding into an already inhomogeneous medium or if feedback has destroyed some of the CO in the shell. If molecular gas is missing, we cannot accurately determine the original molecular mass that was swept up by the superbubble, and thus the total mechanical energy injected (which is needed to determine the cluster properties and feedback mechanism driving the superbubble; see Sect. \ref{sec:derived-props}).

The 88 bubbles with  Closed Ring morphologies is a vast improvement (an order of magnitude) on the number of superbubbles identified in nearby galaxies using molecular gas \citep{bolatto_suppression_2013}. For the rest of this paper, unless stated otherwise, all mentions of superbubbles and their analysis refer to the 88 bubbles labelled as  Closed Rings. We list the bubble positions for the first bubble (i.e. sorted by RA) found in each galaxy in Table \ref{tab:props} and provide an extended version of this table online.\footnote{All tabulated properties can be found at \url{ https://www.canfar.net/storage/vault/list/phangs/RELEASES/Watkins_etal_2023b}}

\section{Sample properties} \label{sec:general-results}

\subsection{Morphology and location}
The 88 superbubbles are not evenly distributed amongst the 18 galaxies. Typically, galaxies with some flocculent spiral structure and weaker bars have more superbubbles. Four galaxies have ten or more bubbles, and at least one bubble was found in every galaxy. Most bubbles show some form of asymmetry in their intensity distributions, such as one side appearing blown out when at the edge of a spiral arm, centrally concentrated knots of CO around the shell, or part of the shell with stronger emission. Very few bubbles have strong ellipticity as expected when tracing younger superbubbles \citep{barnes_phangsjwst_2023}. We also find that around half of the bubbles are located in spiral arm features. When using the environment masks outlined in \cite{querejeta_stellar_2021} (internal release v5) to assign environmental locations to the bubbles, we find around 39--47 are in the spiral arms of the galaxies; the range accounts for the fact some bubbles overlapped with
two environment labels. The next most common environment that bubbles are located in are the inter-arms, with 17--26 bubbles; 8--11 are in the bars of the galaxies; 1--2 bubbles are located in discs that contained no other strong dynamical feature (such as arms) that we could use to label the environment; and 10--11 did not have an associated environment due to the galaxy type (see Table \ref{tab:props} for exact assignments). We expect that the dense environment of the arms (and therefore the higher density of star formation), in addition to the limited timescales where we can use CO to trace with molecular superbubbles, likely causes us to find bubbles more often within spiral arms. 

\subsection{Sizes}
We find that the mean and median bubble radii are 134~pc and 128~pc, respectively, with a standard deviation spread of 59~pc. In Fig.~\ref{fig:meas-bub_props} we plot the distribution and their statistical measures and provide exact values for their radii in Table~\ref{tab:props}. For the large radii, Fig.~\ref{fig:meas-bub_props} reveals that the radii follow a power law. Using Pareto’s maximum likelihood estimator (MLE), we calculate that the index of the power law is $p=-3.2\pm0.4$. Since the sample size (88) is too small for bootstrapping the uncertainty, the standard deviation is used instead. The power-law index is similar to \HI\ observations in nearby galaxies ($-$2.9 \citealt{bagetakos_fine-scale_2011}), and therefore it reinforces that our sample is representative, although we note that ours is slightly steeper than the numerical prediction ($-2.7$) in \cite{nath_size_2020}. The steeper slope might indicate we find slightly more superbubbles that are smaller, but a bigger sample of molecular superbubbles is needed to investigate this further. 

\begin{figure}
\centering
\includegraphics[width = \columnwidth]{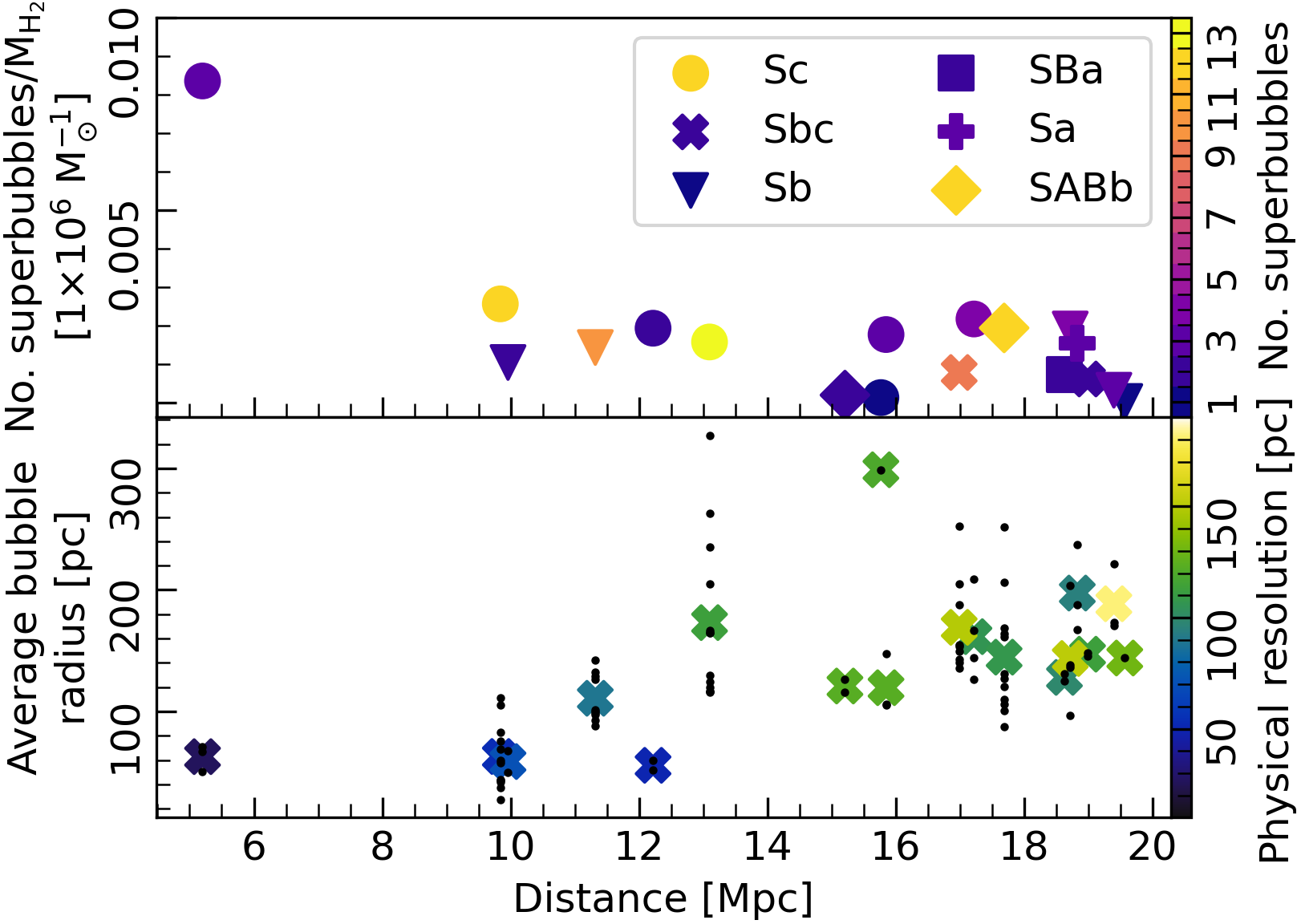}
\caption{Scatter plots indicating the impact of the resolution on the number of superbubbles found per galaxy. \textit{Top}: Number of superbubbles detected per galaxy normalised by the galaxy's molecular mass as a function of distance. The marker shape indicates galaxy type (see legend). The colour of the markers shows the number of bubbles detected. \textit{Bottom}: Mean bubble radius (crosses) per galaxy. The black dots above and below the mean radius shows the individual radii found per galaxy. The colour of the markers shows the physical resolution of the CO data.}
\label{fig:complete}
\end{figure}

In total, the bubble radii are in the range 30--330~pc, which roughly corresponds to our resolution limit; the smallest bubble we measure is located in one of the closest galaxies (Bubble 7 in NGC~628), and therefore we expect to detect a greater number of smaller bubbles with higher-resolution data \citep{watkins_phangsjwst_2023}. To test the distance completeness, we plot the number of superbubbles detected per galaxy divided by the total molecular mass of the galaxy, against the distance to that galaxy in the top panel of Fig.~\ref{fig:complete}. We find that fewer bubbles are detected as the distance increases, indicating that resolution reduces the number of superbubbles found.

To confirm that the resolution limits the number of bubbles detected, we also plot the average bubble radius per galaxy against the distance to that galaxy in the bottom panel of Fig.~\ref{fig:complete}, and colour them using the physical resolution of that galaxy. Again, the plot shows that the average bubble radius increases with the distance, and that the majority of averaged radii are around 1.5 times larger than the physical resolution, highlighting the need for high-resolution observations when conducting a survey of molecular superbubbles. Interestingly, we find that for the closest galaxy, NGC~5068, the average radius (60~pc) is double the physical resolution (30~pc), which might suggest that 30~pc reaches a physical turnover, though with only three superbubbles in NGC~5068 it is hard to tell. \cite{watkins_phangsjwst_2023} also find a turnover at 30~pc in NGC~628, but instead attribute the turnover to the completeness limit. Clearly, more data at higher resolutions are needed to confirm which interpretation is more appropriate. 

While we have the largest sample of molecular superbubbles detected in nearby galaxies, Figs.~\ref{fig:meas-bub_props} and \ref{fig:complete} indicate that our sample is likely biased towards more extreme superbubbles with larger cluster masses (where superbubbles can grow to larger sizes over shorter timescales, whilst still retaining a significant amount of gas in a molecular shell). We discuss the nature of these superbubbles and compare the number we detect to an estimated number of similar superbubbles in Sect.~\ref{sec:bub-loc}.

\subsection{Shell mass and expansion velocity} \label{sec:mass-velo}
Measuring a representative shell mass ($M_\text{sh}$) has additional challenges: the shells are not resolved, the shells are not always perfectly isolated, and some have asymmetric features. Given these circumstances, we decided to measure the bubble masses using a fixed percentage of the bubble radius. The first value we tried was based on the FWHM of the average, normalised $T_\text{peak}$ intensity profile of all the bubbles in Fig.~\ref{fig:profile} (i.e. the normalised radius where the normalised intensity dropped by 50\%). This occurred at 1.5$R$, where $R$ is the bubble radius. However, when plotting an aperture at 1.5$R$ for each bubble, we find that it slightly underestimates the shell size for smaller bubbles. Since the large bubbles tended to appear more isolated (i.e. have less emission around them), we decided to increase the radius slightly to 1.6$R$ when measuring masses, which is still within the statistical spread, defined using the standard deviation shown in Fig.~\ref{fig:profile}. At 1.6$R$ the statistical spread is around 15\%; therefore, we use this as an additional uncertainty when calculating the mass of the bubbles. We converted the CO luminosity into mass using a CO-to-\h\ factor ($\alpha_\text{CO}$) from \cite{sun_molecular_2020} which accounts for the mean variation of $\alpha_\text{CO}$ with galactocentric radius using a mass-metallicity relationship. The mean and median molecular mass of the bubbles are 19\e$^6$~\msun\ and 13\e$^6$~\msun,\ respectively, with a standard deviation of 18\e$^6$~\msun. The large standard deviation reflects the two order of magnitude spread in masses (0.2\e$^6$~\msun\ to 67\e$^6$~\msun). While the median masses are high, we show that these superbubbles are driven by large stellar populations in the next section (Sect.~\ref{sec:hst-results}), and therefore that higher shell masses are expected. In Fig.~\ref{fig:meas-bub_props} we show the mass distribution and their average values. Altogether, we find 2\% of all the molecular gas found within the 18 galaxies is contained within our sample of superbubbles (6\% including all 325 superbubbles).

Realistic expansion models show that the 3D geometry of superbubbles are highly asymmetric and elongated perpendicular to the galaxy disc, due to lower densities and pressures at larger scale heights or away from the central star-forming disc \citep{baumgartner_superbubble_2013}. This results in a peanut-like bubble morphology where the higher-density gas near the disc confines the shell, and large lobes expand perpendicular to the disc (i.e. they do not expand spherically). However, we expect that the distorted geometry of the bubbles becomes pronounced only after their vertical extent grows beyond three times the scale height of the galaxy, which is $\sim300$~pc for molecular gas and $\sim$1~kpc for the ionised gas \citep{baumgartner_superbubble_2013}. Given that the average bubble radius is $\sim$130~pc for our sample, we can ignore the impact of complex geometries and opt for a spherical approximation for the bubble expansion.

\begin{figure}
\centering
\includegraphics[width = \columnwidth]{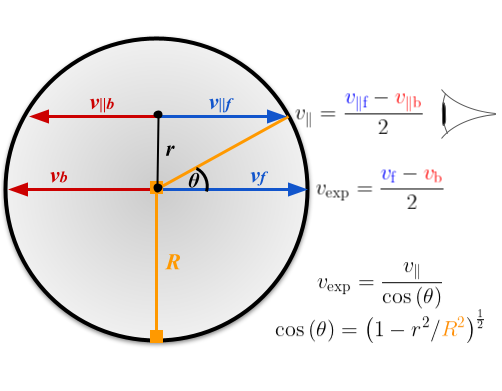}
\caption{Illustration of how expansion velocity is calculated.}
\label{fig:Vexp}
\end{figure}

For a spherical model, the expansion velocity is half the difference between the velocity measured at the back (receding: red-shifted) and front (approaching: blue-shifted) sides of the bubble at the bubble centre. This idealised measurement is not always possible, due to the CO sensitivity towards the centre of some bubbles. Therefore, if measured away from the centre, we only view a component of the expansion velocity (along the line of sight) and have to multiply the denominator by a cosine function (see Fig.~\ref{fig:Vexp}). Altogether, the equation to measure expansion velocity is given by
\begin{equation}
    v_\text{exp} = \cfrac{v_{\parallel\text{b}}- v_{\parallel\text{f}}}{2\left(1-r^2/R^2\right)^{\frac{1}{2}}},
    \label{eq:vexp}
\end{equation}
where $v_\text{exp}$ is the expansion velocity, and $v_{\parallel\text{f}}$ and $v_{\parallel\text{b}}$ are respectively the velocities measured at the front (approaching) and back (receding) of the bubble at the position $r$, where $r$ is the distance from the bubble centre and $R$ is the bubble radius.

To measure the expansion velocities, we make a horizontal and vertical PV diagram of each bubble. In PV space the bubbles also appear as a hole where the emission splits into two separate velocities and so we measure their red-shifted (back) and blue-shifted (front) velocity by hand (see Fig.~\ref{fig:method}, panel 4). The two expansion velocities are then averaged together. In general, the two expansion velocity measurements are similar, with an average difference of 1.7~\kms\ between the two. With channel widths of 2.5~\kms, we estimate the total measurement uncertainty is 3.0~\kms\ for each bubble. We note that these errors are smaller than the (spherical) model uncertainty. If emission is visible towards the centre, we measure their velocities there, otherwise we measure them closer to the edge. Thirty bubbles in the sample (one-third), have at least one measurement taken away from the centre. Additionally, a small fraction only exhibit velocity expansion on one side of the bubble in PV space (such as Bubble 36 shown in Fig.~\ref{fig:method}, panel 4). For these, we measure the expansion velocity using the difference between the velocity in the centre of the cavity in PV space and the bubble edge that we can see. When we do this, Eq.~\ref{eq:vexp} is multiplied by two. The mean and median expansion velocity of the sample are 9.8~\kms\ and 8.9~\kms,\ respectively, with a standard deviation of 4.3~\kms, which is fairly typical for expanding superbubbles, and matches our expectations given the sound speed of ionised gas \citep{krumholz_dynamics_2009} and the typical expansion rates of bubbles \citep{rahner_winds_2017}. Furthermore, these values are consistent with the expansion velocities reported in \cite{kruijssen_fast_2019}, \cite{chevance_lifecycle_2020}, and \cite{kim_environmental_2022}, which were calculated independently using a statistically derived feedback timescale. In Fig.~\ref{fig:meas-bub_props} we plot their velocity distributions.

Finally, using the mass and expansion velocity, we calculate the kinetic energy ($\frac{1}{2}M_\text{sh}v_\text{exp}^2$) and find that the mean and median are 3.2\e$^{52}$~erg and 0.8\e$^{52}$~erg, respectively, with a standard deviation of 6.1\e$^{52}$~erg. Again, the large statistical spread reflects the orders of magnitude spread in mass values.

\begin{figure}
    \centering
    \includegraphics[width =\columnwidth]{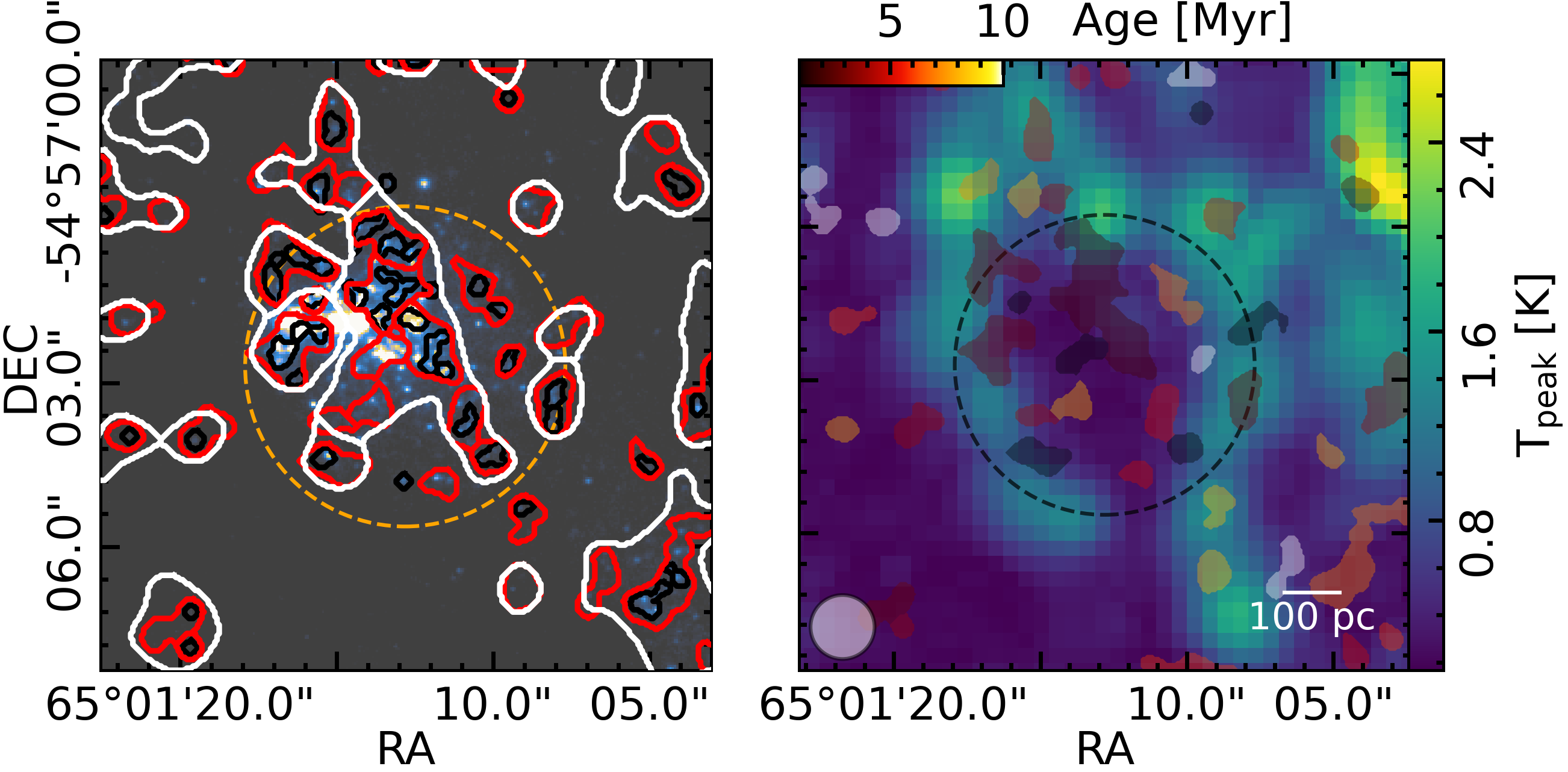}
    \caption{Typical stellar associations, their overlap with superbubble features, and their appearance using bubble 36 from NGC~1566 as an example (see blue box in Fig.~\ref{fig:method}, panel 2) \textit{Left}: HST \textit{B}-band image with 16~pc (black contours), 32~pc (red), and 64~pc (white) association catalogues overlaid. \textit{Right}: CO $T_\text{peak}$ map with the 32~pc associations overlaid coloured by their age. The filled translucent circle indicates the beam and the white line indicates a physical scale of 100~pc. The dashed orange and black circles show the bubble radius.}
    \label{fig:assoc_example}
\end{figure}

\begin{figure*}
    \centering
    \includegraphics[width = \textwidth]{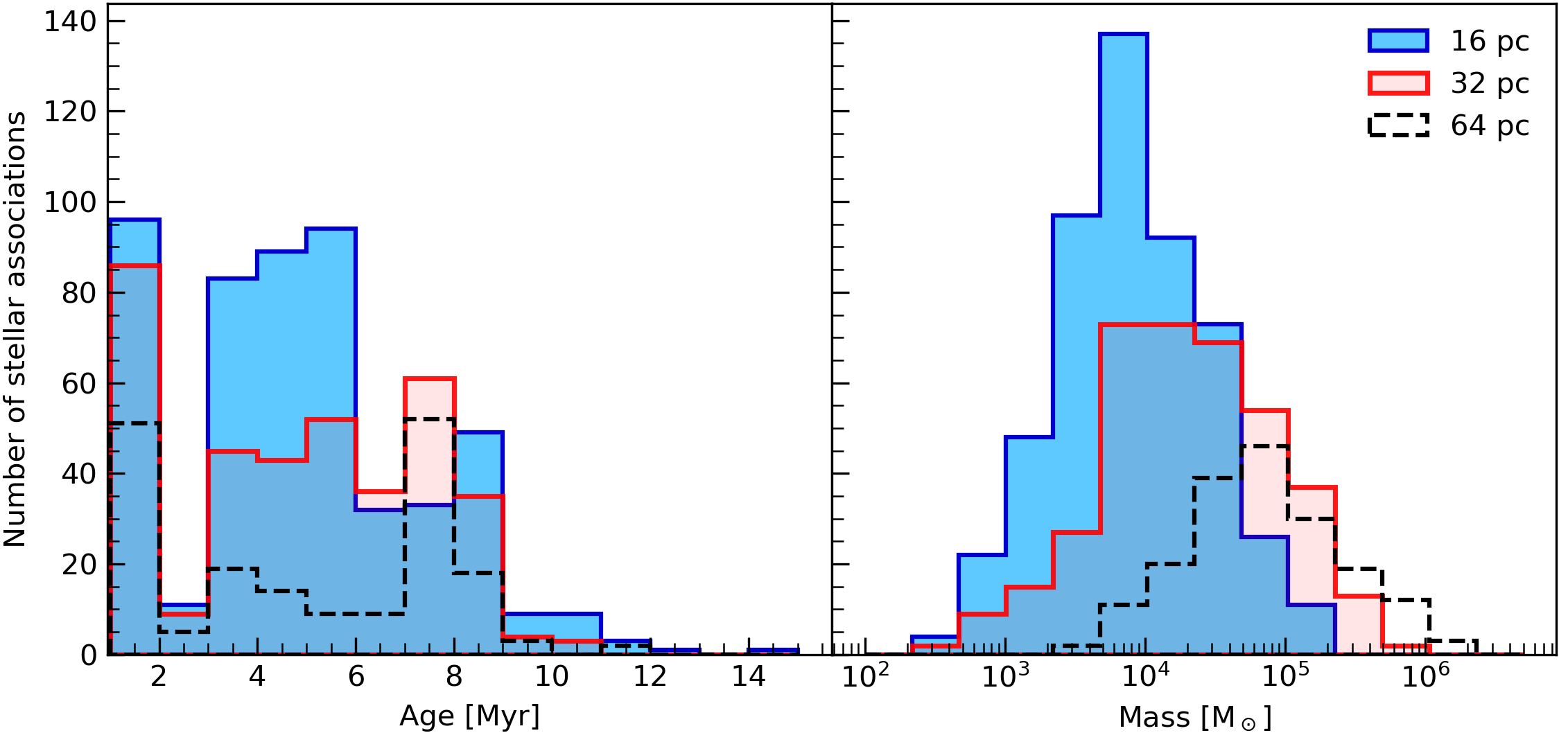}
    \caption{Distribution of the ages (left) and masses (right) for all stellar associations measured within the bubble radius at the spatial scales 16~pc (blue), 32~pc (translucent red), and 64~pc (dashed black).}
    \label{fig:hst_hist}
\end{figure*}

\subsection{Stellar association properties} \label{sec:hst-results}
To estimate ages and masses of the central stellar components associated with the superbubble expansion with HST we had two potential choices: 1) the compact star cluster catalogues that include symmetric (class 1) and asymmetric (class 2), slightly resolved star clusters (see e.g. \citealt{whitmore_star_2021} and \citealt{thilker_phangs-hst_2021}); 2) the multi-scale stellar associations catalogues that include resolved groupings of individual stars using three different scales (16~pc, 32~pc, and 64~pc) to identify the associations (see \citealt{larson_multi-scale_2022}). In this work we only use the multi-scale stellar associations since they include a larger fraction of the stars. Ages determined from the compact clusters and from the related multi-scale stellar associations that often surround the compact clusters are generally in very good agreement.

In Fig.~\ref{fig:assoc_example} we illustrate how the stellar associations appear in relation to the superbubbles. All associations that fall within the bubble radius are assigned to the superbubble. If only a fraction of the association falls within the bubble radius, we multiply its mass by the overlapping fraction. In total, we find that a median of 5, 3, and 1 associations overlap with the superbubble radius at scales 16, 32, and 64~pc, respectively: as the spatial scale increases, more associations are merged, which reduces the number of individual associations identified. We plot their unaltered mass and age distributions at each physical scale in Fig.~\ref{fig:hst_hist}.

The median and mean masses of these distributions are 0.7\e$^{5}$~\msun\ and 1.6\e$^{5}$~\msun, 1.9\e$^{5}$~\msun\ and 49\e$^{5}$~\msun, 7.3\e$^{5}$~\msun,\ and 16.3\e$^{5}$~\msun\ for the 16, 32, and 64~pc catalogues, respectively; the median and mean ages are 4.0~Myr and 4.4~Myr, 5.0~Myr and 4.4~Myr, and 5.0~Myr and 4.6~Myr for the 16, 32, and 64~pc catalogues, respectively. The average age and mass increase with the size scale, which is expected given that each physical scale encompasses a larger contour of stars. We note here that we found a small number of outliers with ages $>200$~Myr, which either biased the average age or raised the total mass well beyond the average total mass of the sample. On average, the oldest association should be $<$15~Myr old considering the average radii and expansion velocities of the superbubbles are 134~pc and 9.8~\kms; therefore, we excluded all associations with ages above 15~Myr.

To assign a single age and mass for each superbubble, we use a mass-weighted mean for the age, and a sum to find the total mass, where the mass is adjusted by the fraction of area that overlaps with the superbubble. For the weighted mean age per bubble, we derive their uncertainty by combining their catalogued uncertainty with the mass weights using the equation
\begin{equation}
    \sigma_w = \sqrt{\cfrac{\Sigma w_i}{\Sigma w_i/\sigma_i^2 }}
    \label{eq:weighted_sigma},
\end{equation}
where $\sigma_w$ is the uncertainty, $w_i$ are the normalised mass weights, and $\sigma_i$ are the individual association age uncertainties. If any age was listed with an uncertainty of 0~Myr, we replaced it with a value of 0.5~Myr since the age would still round to an integer value using this uncertainty. Finally, for the total mass per bubble, we used the root mean squared (rms) to express the uncertainty. We list the averaged ages and masses, and the number of associations for the first superbubble per galaxy (ordered by RA) at each physical scale in Table \ref{tab:hst_props}, and provide the full version of this table online. \footnote{All tabulated properties can be found at \url{ https://www.canfar.net/storage/vault/list/phangs/RELEASES/Watkins_etal_2023b}} For the rest of the work, any mention of HST values (such as ages) specifically refer to these per-bubble averaged values for age and mass unless otherwise stated. 

\section{Constraining cluster masses and dynamical age using CO} \label{sec:derived-props}

For a self-similar thin-shell solution, the equations describing bubble motion are related by a scaling relation at any two given times. For this solution the age of the bubble, $t_\text{dyn}$, is given by
\begin{equation}
    t_\text{dyn} = \eta\frac{R_\text{s}}{v_\text{s}},
    \label{eq:tage}
\end{equation}
where $R_\text{s}$ is the shock radius and $v_\text{s}$ is the shock velocity. Typically, we assume the shock radius equals the shell radius. The scaling constant $\eta$ determines the rate and size of the expansion and describes whether the bubble is driven by a continual injection of energy (e.g. winds) or a blast wave (SN), and whether the bubble interior and/or exterior shock are non-radiative or radiative \citep{ostriker_astrophysical_1988}. Therefore, correctly determining what drives the bubble expansion has a large impact on ages derived for superbubbles. 

\begin{figure}
    \centering
    \includegraphics[width =\columnwidth]{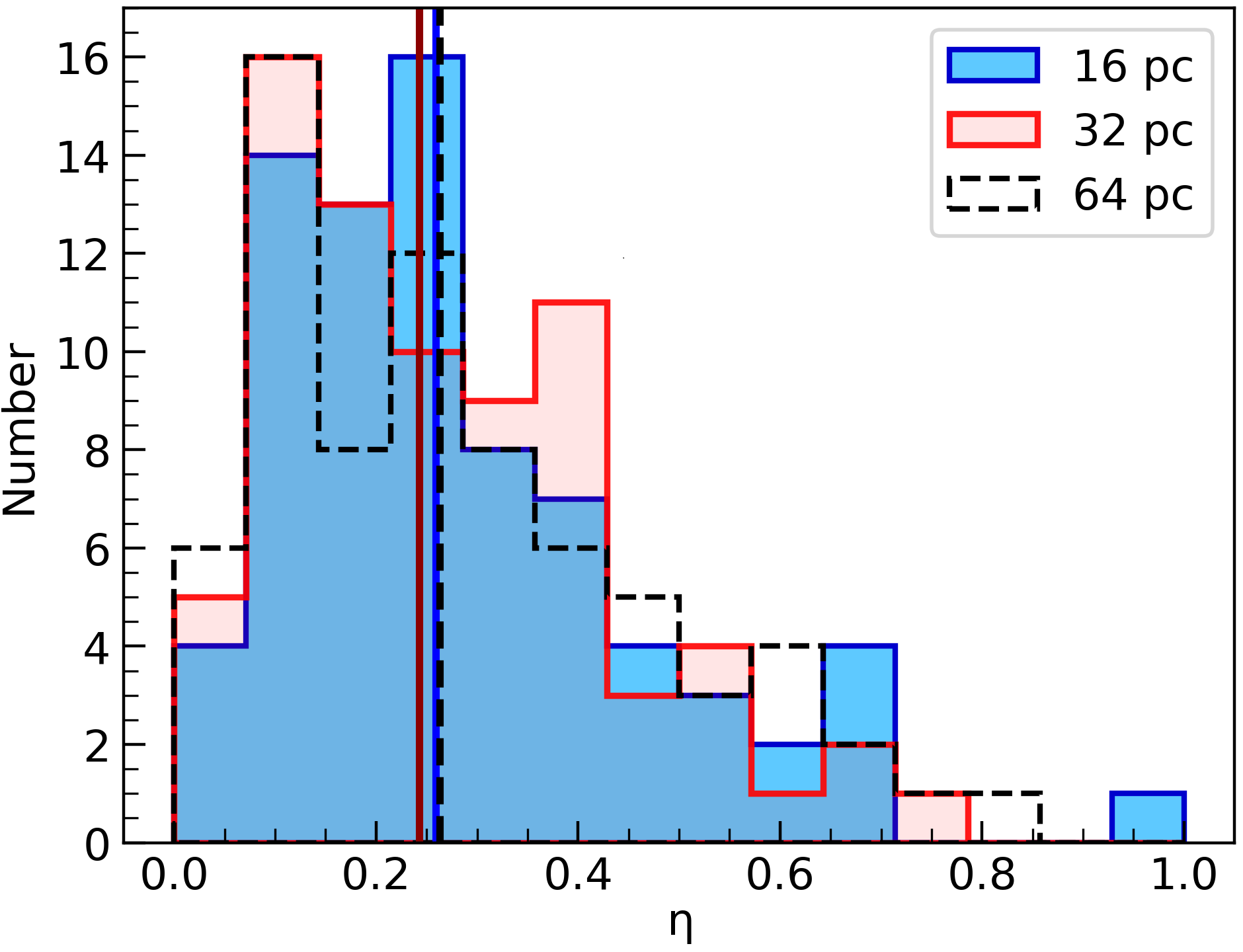}
    \caption{Distribution of $\eta$ values using mass-weighted average HST association age for associations found within each bubble at the spatial scales 16~pc (blue), 32~pc (translucent red) and 64~pc (dashed black). The same coloured vertical lines demonstrate the median of each distribution.}
    \label{fig:eta}
\end{figure}

If we assume the usual jump conditions (i.e. mass, momentum, and energy conservation) for a thin shell where the pre-shock gas is at rest with respect to the bubble centre, the average shock velocity and post-shock velocity can be approximated as
\begin{equation}
    \overline{v} \simeq v_\text{ps} = \frac{2}{1+\gamma}v_\text{s},
    \label{eq:vs}
\end{equation}
where $\overline{v}$ is the average velocity, $v_\text{ps}$ is the post-shock velocity, and $\gamma$ is the adiabatic constant that depends on whether energy is able to radiate (and if the sound speed of the gas is less than the expansion velocity of the bubble, which for CO is almost certainly the case). For radiative models $\gamma=1$. Importantly, we are assuming $v_\text{exp}=v_\text{ps}$. The final two assumptions we make are that the dynamical age of the bubbles measured via their expansion velocities equals the age of the stellar population driving the bubble, which is less accurate for superbubbles that are primarily driven by SNe due to the delay after the first SN explodes. We also assume that the mechanism currently acting dominates the expansion, allowing us to measure a single value for $\eta$, when deriving the bubble properties, though we note that in reality multiple mechanisms contribute towards the bubble expansion at different stages over the lifetime of the bubbles.

Usually, given that emission tracers (such as CO) only provide a radius and a velocity, $\eta$ has to be assumed to derive the superbubble age, which would add an uncertainty in our ages of at least a factor of two. However, by using the average ages for 16, 32, and 64~pc stellar association catalogues found within the bubbles (and the radii and expansion velocities), we can rearrange Eq.~\ref{eq:tage} to measure $\eta$. Doing this, we plot the distribution of $\eta$ values, along with their median values, in Fig.~\ref{fig:eta}. We find the median $\eta$ values for each stellar association scale are $0.26^{+0.25}_{-0.14}$, $0.24^{+0.26}_{-0.14}$, and $0.26^{+0.36}_{-0.17}$. The closest model to these are SN driven snowploughs (momentum driven: $\eta=1/4$, pressure driven $\eta=2/7$), although the tail of the distribution is also consistent with continual radiative energy injection models, which have $\eta = 1/2$. 

\begin{figure}
    \centering
    \includegraphics[width =\columnwidth]{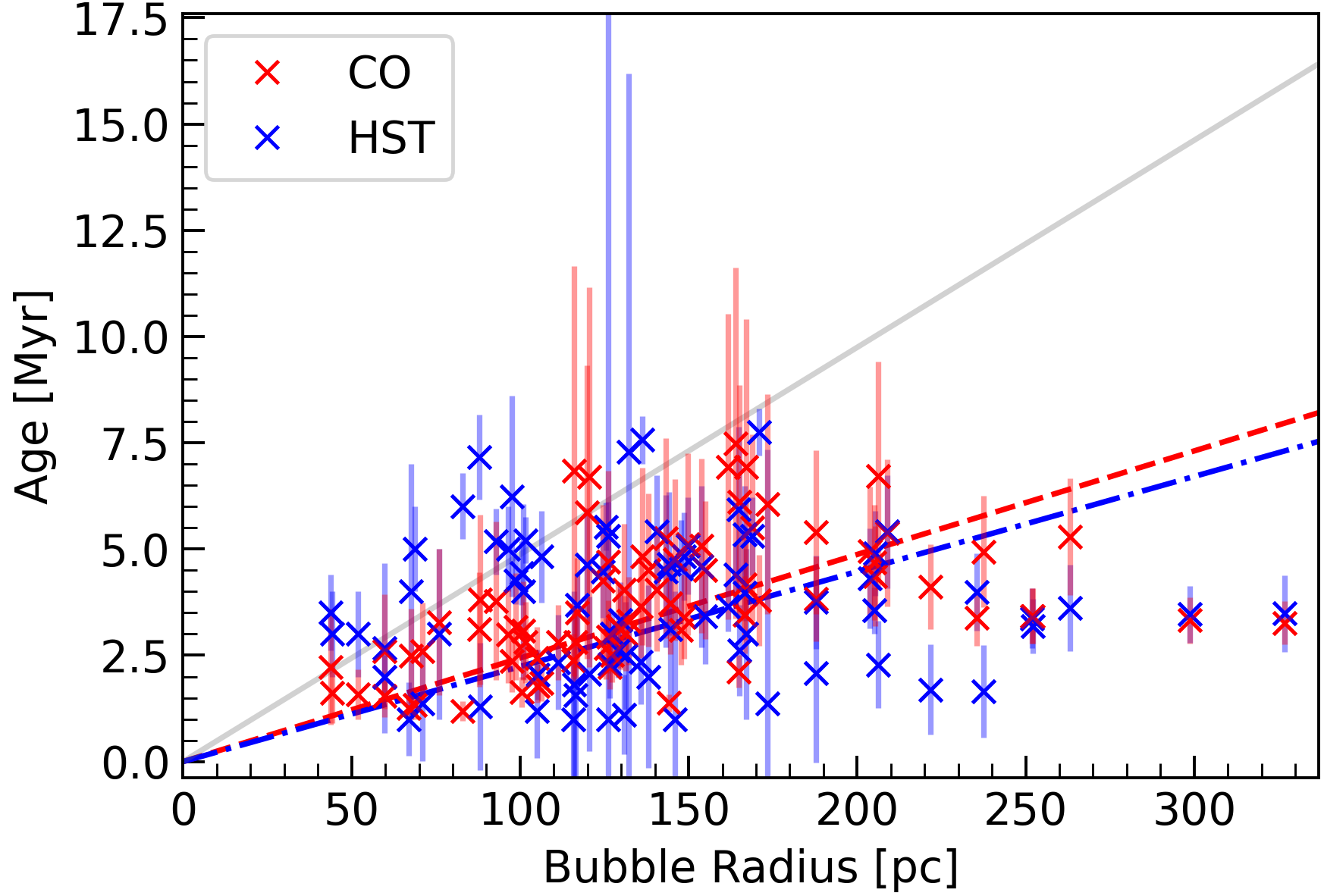}
    \caption{Age vs radius for each superbubble. The blue crosses with vertical lines show the average ages of the stellar associations powering the superbubbles for the 16~pc stellar associations (HST ages) and their uncertainties. The dot-dashed blue line is a linear fit to the blue crosses. The red crosses with vertical lines show the dynamical derived CO ages of the superbubbles and their uncertainties, assuming they follow a self-similar solution driven by SNe ($\eta=1/4$), which is the closest physical model that matches the HST ages (see Fig.~\ref{fig:eta}). The dashed red line is a linear fit to the red crosses. The solid grey line shows a linear fit to the CO-derived values if we used a self-similar solution describing continual radiative energy injection models ($\eta=1/2$).}
    \label{fig:age-rad}
\end{figure}

To investigate further, we plot the mean weighted age of the 16~pc associations identified within the superbubble versus their measured radii, and the CO derived values assuming they follow $\eta$=1/4 in Fig.~\ref{fig:age-rad}. We also show a linear fit to both the CO and the HST to better show the relationship between them.
We see that HST ages and the dynamical age derived using CO are well matched. If we instead calculate the dynamical age using $\eta$=1/2, and show the linear fit to these new ages as a grey line in Fig.~\ref{fig:age-rad}, we can see that some of the HST ages agree better with $\eta$=1/2. 

This provides two important results. First, {the average superbubble traced with CO at 100~pc scales expands with a power law of $\sim$0.26. The closest models to $\eta=0.26$ using theoretical self-similar solutions are SN blast wave models. Second, if we adopt this model, then using CO we are able to constrain the age of the stars driving superbubbles similar to \HI\ studies. This work therefore provides us a new way to measure the impact of feedback using only CO, and also informs theoretical and numerical studies constraining superbubble feedback.}

\begin{figure}
    \centering
    \includegraphics[width =\columnwidth]{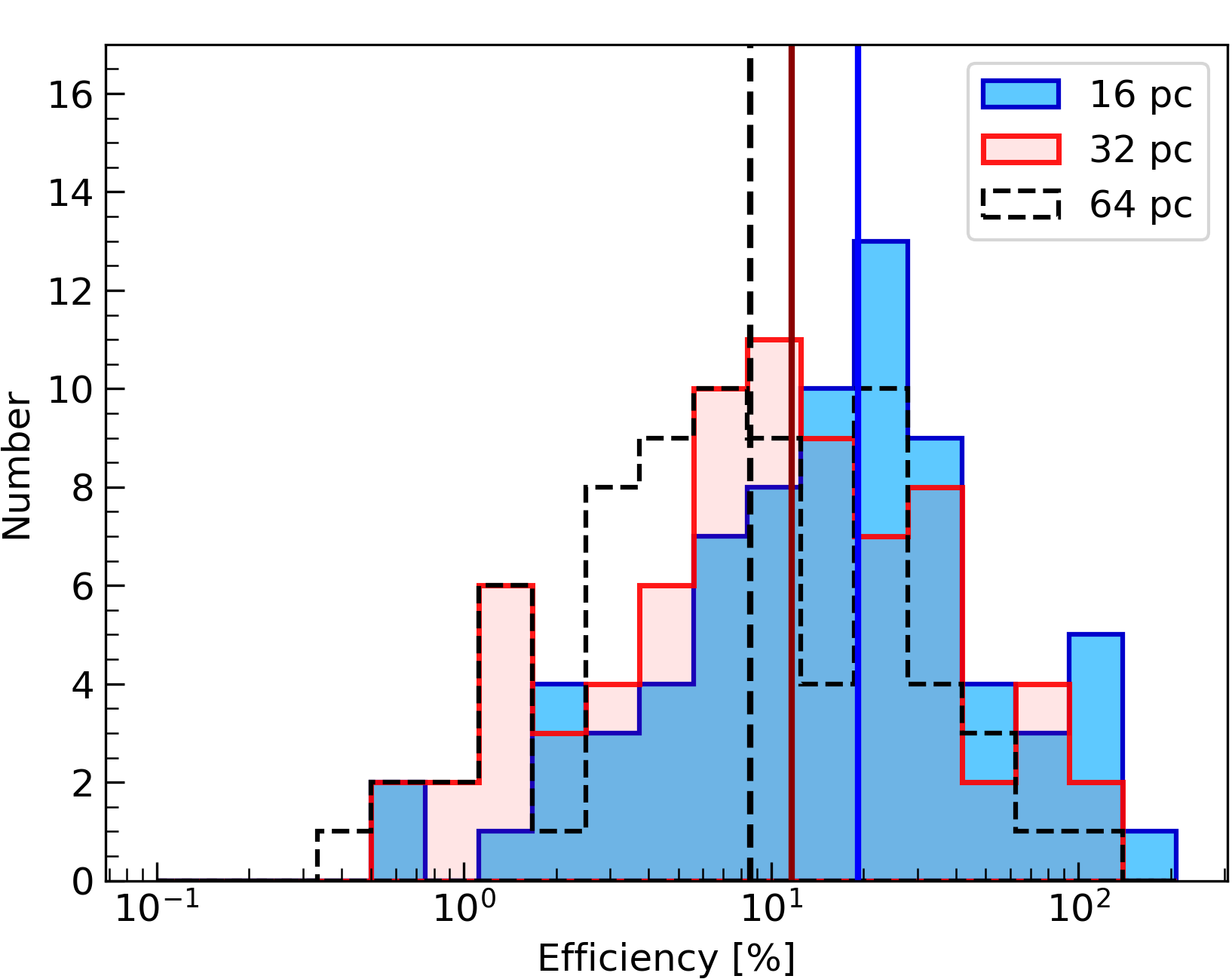}
    \caption{Distribution of efficiency values needed for CO derived cluster mass to match the total HST stellar association mass within each bubble at the spatial scales 16~pc (blue), 32~pc (translucent red), and 64~pc (dashed black). The same coloured vertical lines demonstrate the median of each distribution.}
    \label{fig:effic}
\end{figure}

With the age defined, we model the mass of stars needed to power the measured bubble expansion by comparing the kinetic energy injected in the superbubbles shells over their lifetime (using CO) to the total mechanical power outputted by its stellar population. Using Starburst99 models \citep{leitherer_starburst99_1999}, a stellar population with a 1\e$^6$~\msun\ at solar metallicity following a fully sampled initial mass function \citep{chabrier_galactic_2003} has a mechanical luminosity of $\sim$1\e$^{40}$~erg~s$^{-1}$ (SNe and wind combined) for ages $\lesssim4$ Myr, after which Wolf-Rayet winds increase the rate at which energy is injected. The power injected into the superbubble shell is equal to its kinetic energy multiplied by its age. Therefore, the difference between the two directly estimates the cluster mass. However, the powering cluster does not impart all of its energy into the surroundings, and without a baseline measurement we would have to use theoretical models to predict the efficiency and derive the stellar mass. Therefore, the equation for the cluster mass ($M_\text{cl}$) is
\begin{equation}
    M_\text{cl} = \cfrac{M_\text{sh}v_\text{exp}^2t_\text{dyn}}{2L\epsilon}
    \label{eq:M_CO}
,\end{equation}
where $L$ is the average mechanical luminosity and $\epsilon$ is the injection efficiency. Since we know the stellar mass driving the bubbles using the HST stellar association catalogues, we can constrain what values of $\epsilon$ are needed to match the CO-derived stellar mass to the stellar association mass by dividing the two in Fig.~\ref{fig:effic}.

Rearranging Eq.~\ref{eq:M_CO} for $\epsilon$, we measure average efficiencies of 9--19\%. Specifically for 16, 32, and 64~pc, we measure 19$^{+60}_{-14}$\%, 12$^{+27}_{-9}$\%, and 9$^{+36}_{-6}$\% respectively. In doing this calculation, we use a simple constant value for the mechanical luminosity per unit stellar mass (1\e$^{34}$~erg~s$^{-1}$~M$_{\odot}^{-1}$) even for superbubbles with ages $>4$~Myr since we found that the average efficiency in Fig.~\ref{fig:effic} was unaffected when integrating the continuous injection of energy. We also note that for the smaller stellar populations with $<10^4$~\msun, stochastic sampling of the IMF can affect the mechanical energy by an order of magnitude \citep{da_silva_slugstochastically_2012}. Since almost all of the total HST masses within the superbubbles surpass $10^4$~\msun, we assume that stochasticity has a small impact on the estimate for the mechanical luminosity.

\begin{figure}
    \centering
    \includegraphics[width = \columnwidth]{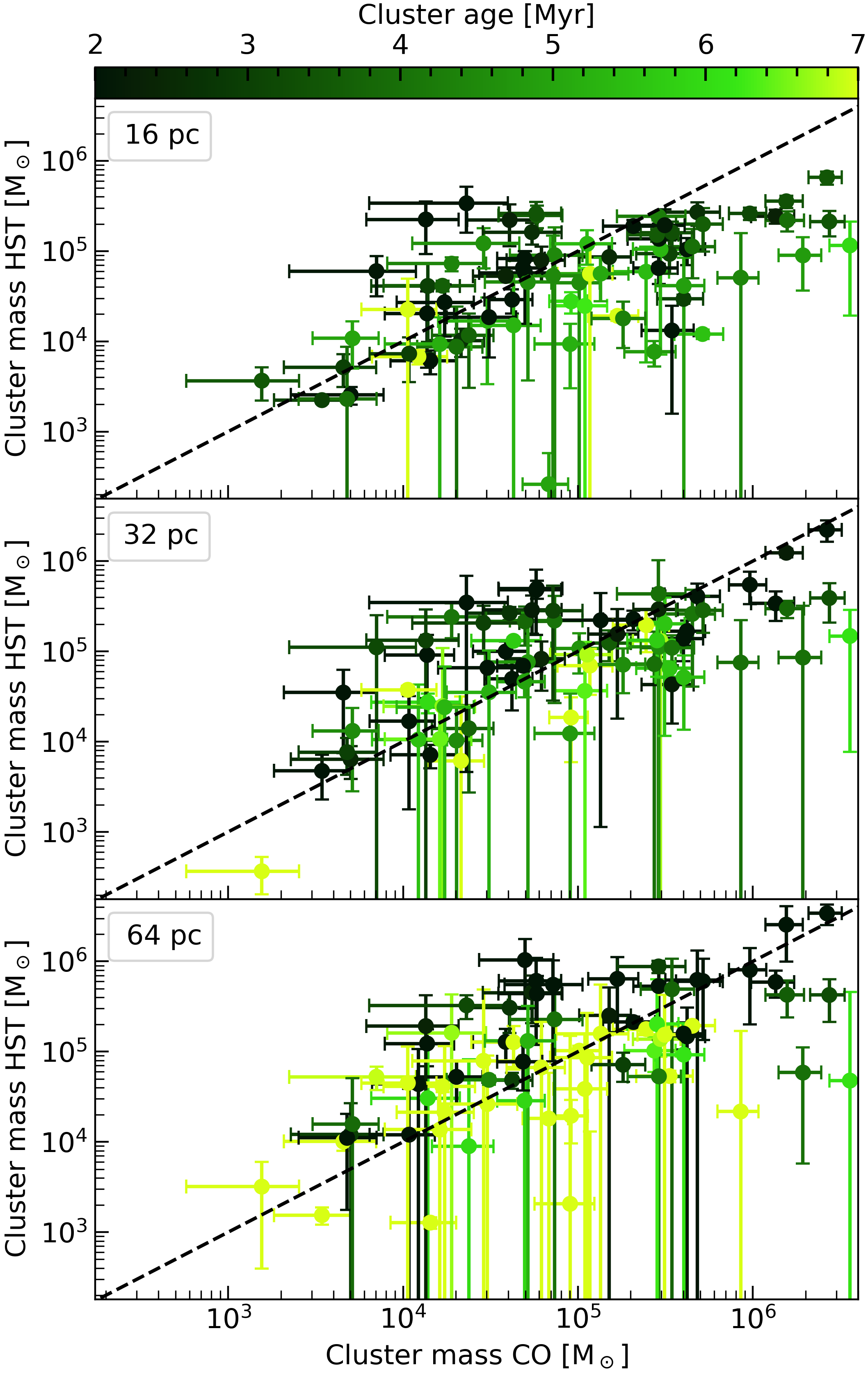}
    \caption{Total HST association mass vs dynamically derived CO stellar mass for each stellar association catalogue. \textit{Top}, \textit{middle}, and \textit{bottom} panels show the masses from the 16~pc, 32~pc, and 64~pc multi-scale association catalogues, respectively. The dashed black line shows the one-to-one line for an $\epsilon$ of 10~\%. The markers are colour-coded by their age.}
    \label{fig:mass-sc}
\end{figure}

In Fig.~\ref{fig:mass-sc} we show the CO-derived mass using an injection efficiency of 10\% for each stellar association scale, where 10\% is between the values found for the larger stellar association scales. We note here that since we know that the 16~pc scale stellar association catalogue underestimates the mass due to the smaller areas they cover \citep{larson_multi-scale_2022}, we do not take it into consideration for the efficiency, but we show it for completeness. The dashed line shows the one-to-one relationship in each panel. We see that a 10\% efficiency fits the observations well. The efficiency is a poorly constrained and highly debated value in theoretical works \citep{cooper_energy_2004,krause_feedback_2013-1,krause_dynamics_2014-1,yadav_how_2017,gupta_lack_2018}; therefore, this result provides a vital observational constraint on energy injection at 100~pc scales. We also note that 10\% was derived for molecular superbubbles with much larger radii ($>$700~pc; \citealt{tsai_molecular_2009-1}), suggesting that the efficiency we find is robust even at larger size scales.

Altogether this section has shown that by assuming $\eta = 1/4$ and an injection efficiency of 10\%, the CO is able to reproduce properties of the driven stellar sources. While these model dependent quantities are quantitatively constrained by the HST stellar associations, the fact that this still results in a one-to-one relation for both the ages and the mass is a remarkable result, and future work using CO to characterise superbubbles can now be undertaken using the relations presented here to constrain the ages and masses of the stellar populations.

\section{Discussion} \label{sec:discuss}
For this discussion section we focus on understanding the implications of finding molecular gas in superbubble shells. More specifically, we discuss whether the number of bubbles we find is representative, what mechanisms lead to molecular superbubbles, the origin of the molecular gas in the shells, and the role of feedback in altering star formation histories within superbubbles.

\subsection{Number of molecular superbubbles} \label{sec:bub-loc}

We find a total of 88 near-perfect examples of superbubbles, but compared to the $\sim$1700 bubbles found in a single galaxy using JWST (NGC~628; \citealt{watkins_phangsjwst_2023}), 88 seems like a small number. If the majority of bubbles are missing from our catalogue, our results might not be fully representative, which suggests that either CO is not able to trace the full superbubble population, or our bubble identification method failed to detect some candidates. Therefore, we estimate the theoretical number of superbubbles that should be present based on the star formation rate (SFR) of the galaxies, the lifetime of the bubbles, the average cluster that power such a bubble using Eq.~17 in \cite{clarke_galactic_2002}, and the number of bubbles we expect to merge (see \citealt{watkins_phangsjwst_2023}, where it is discussed in detail). Firstly, assuming that the CO-derived cluster masses follow a lognormal-like distribution, the mean CO-derived cluster mass is 4.8\e$^5$~\msun. For the bubble lifetime, while we expect to see bubbles for up to 1--5~Myr, we find the mean bubble lifetime we measure falls in the middle of this range at 3.6~Myr; therefore, we use 3.6~Myr for the bubble lifetime. Finally, the theoretical number of bubbles depends on the SFR of the observable galaxy area. Given the average SFR per galaxy adjusted for the ALMA field of view is 2.5~\msun~yr$^{-1}$ we find that $\sim$340 superbubbles should be identified in the 18 galaxies (19 per galaxy). If 30\% of the bubbles merge \citep{simpson_milky_2012,watkins_phangsjwst_2023}, we estimate $\sim$240 superbubbles should be detected (13 per galaxy). Therefore 88 superbubbles underestimates the superbubble population. However, if we include the extended sample of CO superbubbles, which contains 325, the two estimates are actually comparable to our catalogue. Altogether, the similarity between the estimated number of molecular superbubbles observable at 100~pc scales with an average lifetime of 3.6~Myr to the total number of superbubbles we identified confirms these are real superbubbles and suggests that many objects in the extended catalogue are bona fide superbubbles.

\begin{figure}
    \centering
    \includegraphics[width = \columnwidth]{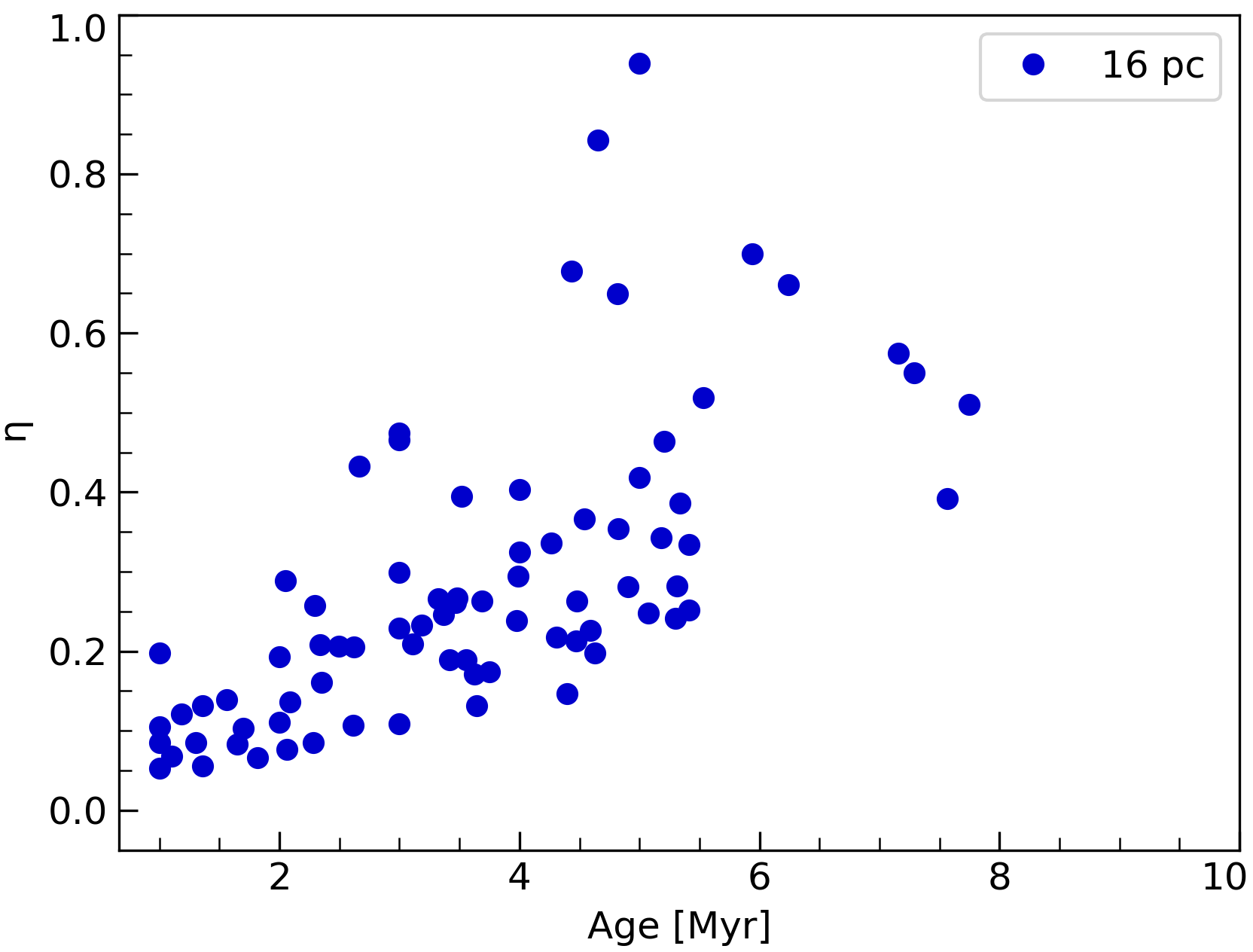}
    \caption{Scaling constant $\eta$ as a function of mass-weighted mean age, using the 16~pc stellar associations found within each superbubble. Values of $\eta$ are typically smaller for younger superbubbles, and larger for older superbubbles. A similar trend is seen when using different stellar association scales.}
    
    \label{fig:eta_vs_age}
\end{figure}

\subsection{Mechanisms driving the superbubbles and their gas expulsion} \label{sec:discuss-feedback}
Superbubbles are a byproduct of feedback acting on the surrounding gas, both pushing it away and destroying the molecules present. By removing gas, feedback interrupts star formation and leaves behind a cavity we can detect. In this section we discuss which feedback mechanisms create the molecular superbubbles in order to better understand the mechanisms that limit the star formation efficiencies (SFEs) within them. We note here that our ability to see superbubble shells in molecular tracers implies that the main impact of feedback in these regions is likely dominated by molecular gas removal rather than molecular gas destruction (i.e. the CO and the stellar associations were co-spatial when the associations initially formed; see the next subsection for a more in-depth discussion).

Assuming superbubbles are well explained by a self-similar expansion model, our results show that SN blast waves dominate superbubble expansion for the majority of molecular superbubbles at 100~pc scales and ages of 1--5 Myr (i.e. $\eta=1/4$ or $\eta=2/7$ are the closest models to $\eta=0.26^{+0.25}_{-0.14}$, see Sec \ref{sec:derived-props}). However, due to the statistical spread we measure for $\eta$, a subset of the superbubbles are consistent with being powered by continuous energy injection ($\eta=1/2$; e.g. stellar winds), and when viewed in Fig.~\ref{fig:age-rad} it manifests as a minority of superbubble ages (traced with HST) following this steeper linear relation. Before the onset of SNe, a continuous injection of energy is expected to be provided by stellar winds, but when multiple SNe occur in homogeneous environments with densities of $\sim$1~cm$^{-3}$ at the same location within a single stellar population, the SNe are predicted to combine and explode in roughly equal time intervals \citep{mac_low_superbubbles_1988} and to become subsonic by the time they reach the superbubble shell. When the shock waves become subsonic, they do not impart an impulse to the shell since they have converted most of their kinetic energy to thermal energy. Therefore, multiple co-spatial SNe can be modelled using a linear injection of energy, which has the self-similar solution of $\eta=1/2$ (similar to stellar winds).

We note that multiple SNe are almost always modelled as such in theory and simulations, but we show that $\eta=1/2$ is only consistent with a minority of our superbubbles. Physically, for this to occur either the self-similar solution is a poor model for the majority of these superbubbles or some of the underlying assumptions that allow one to model SNe as a continuous energy source are not met or are incorrect. For example, when multiple SNe were first modelled in detail, cooling inside the bubble was assumed to be negligible, which meant that multiple shock fronts reach subsonic speeds almost instantly, and so a continuous injection model was the only model that was applicable \citep{mac_low_superbubbles_1988}. However, current simulations show that cooling needs to occur for superbubbles to match observations \citep{lancaster_efficiently_2021}, and so this assumption is not valid for all situations. Since it is beyond the scope of this study to explore how the former can impact our interpretation, we focus on why the underlying assumptions that lead to this model are not met in our sample.

Unlike the usual conditions that superbubbles are modelled to expand into (e.g. ambient densities of $\sim$1~cm$^{-3}$), most of these superbubbles are expanding into a dense molecular medium and are powered by large extended stellar populations. It has been shown that when SNe initially explode and expand in higher densities ($>10$~cm$^{-3}$) and when turbulent mixing is included, the superbubble cools much more efficiently, which decreases the rate that gas from the shell evaporates into the interior \citep{el-badry_evolution_2019}. With less mass, the shock wave from each SNe decelerates at a slower rate as it sweeps up mass and so can remain supersonic for longer. When SN shock waves reach the shell at supersonic speeds, they impart their energy impulsively via a kinetic kick. SNe that occur in different locations due to an extended stellar population ($>$100~pc) while somewhat embedded in dense gas has also been shown to delay the time that SNe interact, resulting in blast waves acting independently, though this is more important at earlier times before a single superbubble cavity has been formed \citep{yadav_how_2017}. Therefore, superbubbles expanding into dense conditions should be driven by impulsive energy injection for a longer time period. We confirm this by plotting the calculated values of $\eta$ from Fig.~\ref{fig:eta} against the derived HST age\footnote{We only show the 16~pc stellar association ages for brevity since we see the exact same trend for the 32 pc and 64 pc scale stellar associations.} in Fig.~\ref{fig:eta_vs_age}, which shows that $\eta$ increases for older stellar populations. We can see the impact this has in Fig.~\ref{fig:age-rad}, where we show that some of the older HST ages (i.e. $>$3.5~Myr, after SNe occur) are better represented by the continuous energy injection model ($\eta=1/2$). Figure \ref{fig:eta_vs_age} has an important implication. Considering that CO only traces the young superbubble population (see previous subsection), we might not be able to trace many superbubbles powered by a pseudo continuous injection of energy via SNe if the superbubbles are delayed by a few million years before they meet the conditions needed to get to this stage of evolution. That is, there is likely an observational bias that decreases the number of bubbles we can see driven by continuous energy injection via SNe that also increases the number of bubbles we can observe driven by SN blast waves.

Considered together, we suggest that radiative winds drive the expansion of molecular superbubbles with pre-SN ages, which initially pushes the molecular gas into a shell, but after SNe occur, they quickly dominate the expansion, leading to superbubbles with kinematics and energetics that are consistent with SN-driven feedback. For most of these superbubbles, the dense ambient medium causes them to be powered impulsively, and only a small fraction appear to be powered by SNe injecting their energy continuously. We expect that as the superbubbles age, they will evolve into a continual energy injection model. To confirm our findings, higher-resolution data sets, and more observations where we can identify molecular superbubbles are needed. The increased number statistics will allow us to populate the parameter space containing older superbubbles, and by resolving the smaller sizes we will be able to trace the driving mechanism earlier in the evolutionary cycle. The younger ages trace will help confirm the timescales on which gas is removed, and ultimately the initial mechanisms that first cleared out the gas within dense molecular superbubbles.

\begin{figure*}
    \centering
    \includegraphics[width = \textwidth]{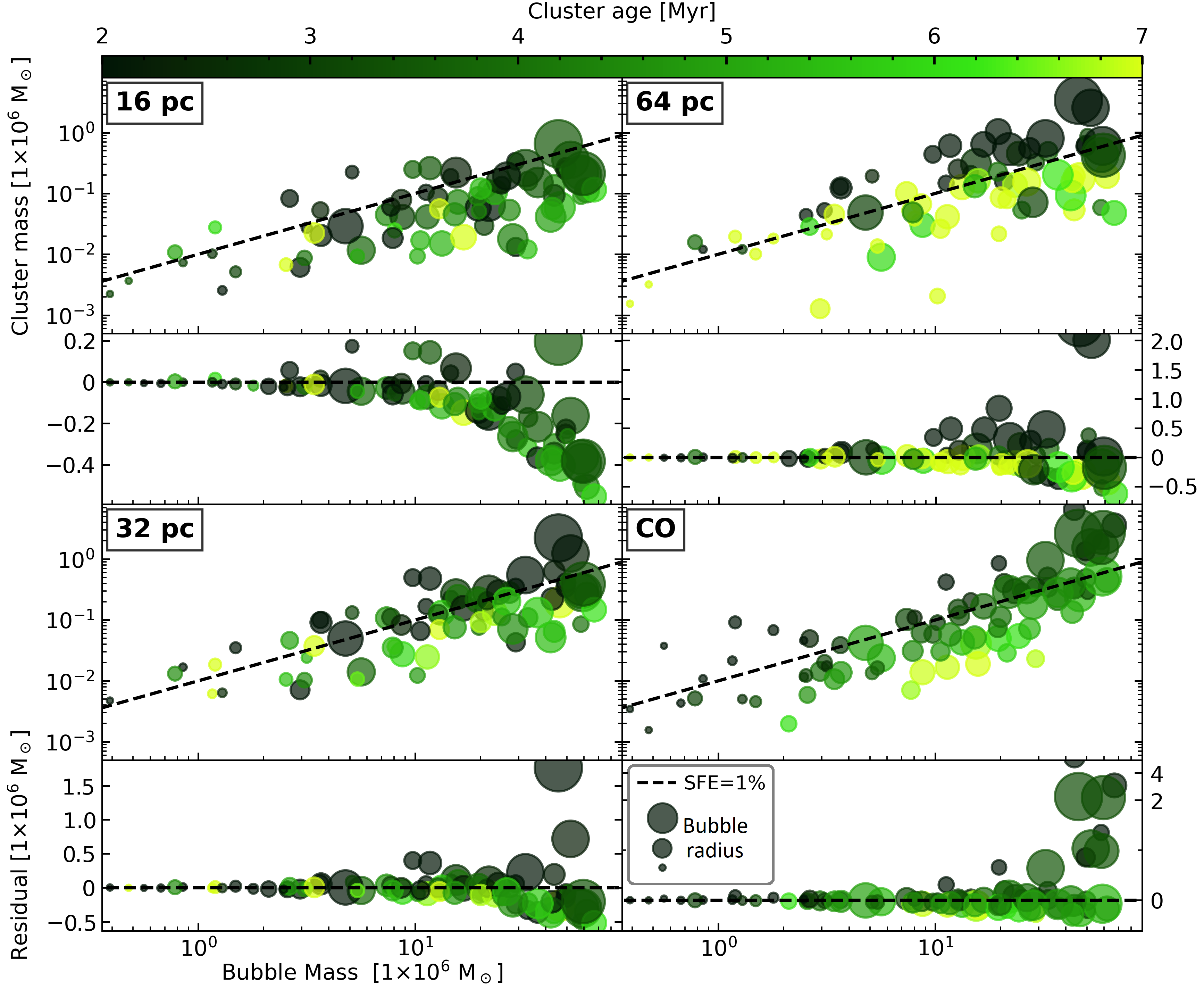}
    \caption{Cluster mass vs shell mass for the three stellar association scales and the CO dynamically derived values. The label in the upper left corner shows which values the subplot represents. The marker colour shows the age for each method and the marker sizes indicate the bubble radius. The dashed black line indicates the values that correspond to a SFE=1\%. For each method, also plotted is the residual each marker is away from equalling a SFE of 1\% directly underneath as a new subplot.}
    \label{fig:swept}
\end{figure*}

\subsection{Origin of mass contained within bubble shell} \label{sec:discuss-sweep}
As superbubbles expand due to feedback, not only can they sweep up material from their natal cloud into a shell, but they can also collect diffuse unrelated gas from their larger surroundings. If the swept-up gas reaches a high enough column density, 
the shell is predicted to fragment and collapses into molecular clumps that can themselves form new stars \citep{elmegreen_sequential_1977,whitworth_preferential_1994,iwasaki_gravitational_2011}. This can even propagate star formation outwards in a chain reaction \citep{inutsuka_formation_2015}. Some authors expect that up to 30\% of stars might form as a result of this mechanism in the Milky Way \citep{thompson_statistics_2012}. Understandably, determining if and how much gas is gathered by superbubbles provides some constraints on the role superbubbles have on star formation rates. 

Usually these calculations are performed using \HI, but \HI\ superbubbles are usually identified further out in galaxies and between spiral arms \citep{bagetakos_fine-scale_2011}. Our sample preferentially identifies molecular superbubbles within spiral structures, providing us with a rare opportunity to investigate if superbubbles gather molecular gas before this gas is destroyed, how much gas they gather, and whether this can propagate star formation. To test this, we assume that all stellar associations within the superbubbles formed with a SFE of 1\% instantaneously \citep{grudic_nature_2019}, and that all the molecular gas from the parent cloud is initially swept up into the bubble shell. In general the average exact percentage is not important, but rather how the relative properties correlate. With a baseline assumption of stars forming at an efficiency of 1\%, superbubbles with a stellar gas-to-mass ratio of 1\% represent regions that have swept up their parent cloud, without sweeping up surrounding ISM, whereas bubbles with efficiencies $<1$\% represent bubbles that have swept up additional material, and so appear less efficient. Furthermore, efficiencies $>1$\% could either represent molecular gas destruction or new star formation in the swept-up shell. Since all of these processes occur in different proportions at a given time, how bubble radii and age change relative to 1\% provides us with a net measure of if and when the bubble is sweeping up mass or destroying mass as a function of bubble expansion and evolution.

Although if the central stellar population powering the bubble is able to lose most of its stellar mass during this time, we are unable to measure shell growth using this method. To check this, we investigate cluster mass loss via stellar ejections. Given that ejected stars can reach velocities on average of 10--20~\kms\ \citep{oh_dynamical_2016}, a star can travel $\sim$60~pc within the average age of the superbubbles (3.6~Myr). This is much less than the average bubble radius (134~pc). More importantly, it takes several million years before a dynamical interaction ejects the star on average, so $\sim$60~pc represents an upper limit, and therefore we do not expect stellar ejections to significantly reduce the observed stellar association mass. Additionally, for a 1\e$^6$~\msun\ cluster, mass-loss rates due to winds and SNe amount to $\sim$3000~\msun~Myr$^{-1}$ \citep{leitherer_starburst99_1999}, and thus we can also ignore this effect. Therefore, we can safely attribute any change in SFE to a change in the molecular mass fraction and correlate them to changes in size and age.

In Fig.~\ref{fig:swept} we plot the change in cluster mass versus the gas mass for each bubble, colour-code each marker by age, and represent the bubble radius using the marker size. We plot these values for the HST-derived association properties at each spatial scale and for the CO-derived properties using Eq.~\ref{eq:M_CO}. In each panel we plot a 1\% efficiency line as a reference. In all four panels we also show the residual distance away from the 1\% line underneath. It is immediately clear in Fig.~\ref{fig:swept} that a SFE of 1\% explains the relationship between stellar mass and shell mass well, suggesting that we were correct to assume that the swept-up mass likely represents the remnant molecular gas that formed the stars powering the superbubbles. Focusing on the swept-up aspect, the figure reveals that small low-mass bubbles reside on the 1\% efficiency line, but as the mass and radius of the bubble increases, the markers fall below this line, indicating an apparent decrease in SFE. Since we have already confirmed that mass loss from the stellar population itself is negligible, and assuming there are no systematic variations in the SFE that depend on mass, the decrease can only occur if the mass within the shell has increased, strongly suggesting that the bubble shells are gaining mass as they expand.
Assuming that most of the gas that is swept up has a density of 50~cm$^{-3}$, which is converted to molecular gas as the superbubble expands in a cylindrical ring given by the molecular scale height (100~pc; \citealt{patra_molecular_2019}), a swept-up shell growing from 100 to 200~pc should gain $\sim30$\e$^6$~\msun\ of gas. By fitting the increase in mass as a function of radius, we find that superbubbles do increase in mass by $\sim30$\e$^6$~\msun\ as their radius increases from 100 to 200~pc. Therefore, the mass growth we see is reasonably well explained via ambient gas that is swept up from the surrounding ISM.

The age trend we observe also agrees with this result, though it is not as obvious. Firstly, we note that while larger bubbles are generally expected to be older, we can also observe large bubbles over a smaller time period if the stellar population powering the superbubble is large; for example, a 1\e$^5$~M$_{\odot}$ cluster can drive a 200~pc bubble in 2.5~Myr \citep{nath_size_2020}. Therefore, we do not necessarily expect to see a positive correlation between age and cluster mass. The correlation we see demonstrates shell growth. The line tracing a SFE of 1\% strongly delineates the old and young superbubbles for the 32 and 64~pc associations and the CO-derived values; the bubbles with older stellar associations all have SFE $<$1\%, while the bubbles with younger stellar associations have SFE $>$1\%. In other words for the same stellar mass, the older ones have significantly more molecular gas in their shells, strongly implying it has been swept up over time. We also see an age gradient for the 16~pc associations, but it is not split by a SFE of 1\%. In Fig.~\ref{fig:mass-sc} we show that the dynamically derived cluster mass exceeds the HST-derived values at 16~pc, and assert it was due to the smaller associations missing some of the HST sources. If the stellar mass is underestimated at 16~pc scales, the SFE will also be underestimated. We note that after accounting for the underestimated mass, the age trend is still weaker at smaller scales, which potentially foreshadows the formation of new young associations. If new, and relatively small, associations form, the mean age should decrease, which would be less apparent at the larger scales where more associations are merged. Finally, we see a few superbubbles with SFE above 1\%. These bubbles have large stellar masses and radii but their shell mass and the age are small. While this could be a genuine increase in SFE, it is far more likely the apparent increase is caused by molecular gas destruction.

These results have strong implications on the survivability of CO, on the impact that feedback has on molecular gas, and therefore on the ability of the gas to continue forming stars. In the next subsection we consider these themes in more depth.

\subsection{Impact on cloud lifetimes by superbubble expansion} \label{sec:discuss-lifetimes}
Recent observations in nearby galaxies and the Milky Way supports a fast but inefficient model for the molecular cloud life cycle. Molecular clouds spend two-thirds of their time starless \citep{battersby_lifetimes_2017, kim_duration_2021} and eventually form stars that quickly remove the molecular gas via pre-SN feedback mechanisms; star formation then halts \citep{grasha_spatial_2019,kruijssen_fast_2019,barnes_which_2020-1,chevance_pre-supernova_2022}. In this model, the star-forming gas is spatially decorrelated from the stars (i.e. the gas is removed or destroyed). In this work we find a similar result: the gas in molecular superbubble shells is likely relocated by a progenitor star-forming region, which blurs the lines on whether the natal molecular cloud has or has not been destroyed. After the onset of feedback we define that a given cloud is destroyed only when its molecular gas has disassociated and now exists in a different phase of the baryonic cycle; when we cannot connect the gas to ionising sources, the gas in the molecular superbubble shells is a deformed version of the molecular gas clouds that originally formed the stellar population driving the bubble. If more stars form in the shell (which we show tentative evidence of in Sect.~\ref{sec:discuss-sweep}), it represents an extension to the cloud lifetime and feedback timescale, and to the star-forming timescale (embedded phase) in the molecular clouds found around superbubbles. Furthermore, when a fraction of these molecular superbubble shells creates another superbubble, some gas can find itself continually exchanged and swept up while remaining molecular \citep{skarbinski_building_2023, jeffreson_clouds_2023}, making it more similar to slow efficient star formation theories, where most of the gas forms stars over a long period of time. 

To investigate how important such star formation is on galactic scales, we measure the fraction of molecular gas found within the superbubbles. We find that 2\% of the total molecular gas within the 18 galaxies is contained within complete superbubble shells on 100~pc scales (see Sect.~\ref{sec:derived-props}); if we use all 325 superbubbles, we find up to 6\%. Therefore, 2--6\% of molecular gas within galaxies can live longer when found in a molecular superbubble on 100~pc scales. This value is small, and only 0.05 and 0.59\% of the galaxy's area contains molecular superbubbles, respectively, for the 88 and 325 molecular superbubbles. However, considering we are missing smaller superbubbles, 2--6\% is a lower limit.

These results imply that 2--6\% of the molecular gas within galaxies is not immediately destroyed by feedback on 100~pc scales. On small scales, and via recent simulations, it has been shown that when dense molecular gas forms, feedback (including pre-SN mechanisms) can do little to destroy it \citep{walch_silcc_2015,haid_silcc-zoom_2019,watkins_feedback_2019-1}. Therefore, we are likely tracing extreme star-forming regions that reached high gas densities before being disrupted by feedback. The high gas densities shield the gas from being ionised, but result in high gas pressures. Combined with the large number of stars that formed, and thus strong wind pressures, the newly formed stellar associations likely remove the molecular gas without destroying it to form a dense shell. It is within these already cavernous voids that SNe acts upon the gas to push it further and even sweep up surrounding diffuse gas that gets in the way. Our results reveal cloud conditions and star formation that result in ineffective feedback and how often this occurs. All together, it provides us with a sample of regions that future studies can use to identify what conditions are needed to limit feedback and predict which young star-forming regions will evolve this way, and if the molecular gas found in these shells are inert or are continuing to form stars.

\section{Conclusions} \label{sec:conclude}
In this work we present the first extensive study of molecular superbubbles at $\sim$100~pc scales in 18 nearby galaxies using PHANGS--ALMA $^{12}$CO (2--1) and PHANGS--HST stellar catalogues.
Utilising co-spatial CO, MUSE, and HST observations, we outlined a superbubble identification and validation method that finds 325 superbubble candidates. Of these, we find a fiducial sample of 88 with convincing superbubble properties (e.g. unbroken shells, measurable expansion velocities). To ensure that our result are robust, we focused only on these 88 superbubbles throughout this work. 

Using CO, we measured observable superbubble properties, such as their radius, mass, and expansion velocity, and using HST catalogues, we also identified the stellar populations (i.e. mass, age) powering the superbubbles. By combining the properties from the two, we directly measured $\eta$ for each superbubble to constrain the feedback mechanism powering the superbubbles. We determine that a SN blast wave model ($\eta=1/4$ or $\eta=2/7$) provides the best match for the stellar ages and radii that we measure for the molecular superbubbles. With the feedback model known, we calculated the dynamical age of the superbubbles using CO derived properties and find they have an average dynamical age of 3.6~Myr and range from 0.7 to 7.5~Myr, matching recently observed CO lifetimes derived using decorrelation methods. The SN model results in a near one-to-one relationship between between the stellar populations and the dynamical age of the superbubble. We also find around one-third of the superbubbles are more consistent with a continuous energy injection model (i.e. stellar winds and a continuous SN injection model). These results confirm that CO is able to trace superbubble properties. The CO properties and stellar populations also allow us to constrain the efficiency with which energy is injected into the gas by comparing the kinetic energy (using CO) to the stellar population present within the superbubbles. We determine SN feedback injects its energy into the gas with an efficiency of $\sim$10\%. 

We also explore the implications of superbubbles retaining molecular gas. We investigate the origin of the mass in the shells and find that the stellar mass associated with the bubble is $\sim$1\% of the molecular shell mass for the smallest bubbles. This suggests that the initial molecular mass within young superbubbles is likely the swept-up remains of the giant molecular complex that formed the stellar cluster, assuming a SFE of 1\%. However, as the bubbles age, we find the apparent SFE falls below 1\%. By ruling out stellar loses as the causes, the drop is best explained if the superbubble sweeps up additional gas as it expands. These results imply that at $\sim$100~pc scales, a minimum of 2--6\% of the total molecular gas within galaxies (i.e. the fraction found within these superbubbles) is not always destroyed by feedback, but is instead relocated, extending the lifetime of the star-forming gas. Finally, we calculated and compared a theoretical estimate of the molecular superbubble population to our catalogue and find that the two are comparable, confirming that our conclusions are based on a representative sample of molecular superbubbles.

In summary, our work demonstrates that molecular superbubbles provide vital information about the feedback physics in nearby galaxies, informing supernovae studies, gas clearing timescales, and the conditions that can lead to gas removal rather than destruction. Furthermore, we can tentatively begin to constrain propagated star formation within superbubble shells. All of these processes lead us one step closer to understanding star formation histories within galaxies, and therefore what sets observed star formation rates.

\begin{table*}[]
    \caption{Superbubble properties ordered by RA derived using CO. The final column contains the environmental location of the bubbles \citep{querejeta_stellar_2021}. Only the first superbubble from each galaxy is listed.}
\resizebox{\textwidth}{!}{
\begin{tabular}{rrrrrrrrrrr}
Bubble & Galaxy  & RA                   & Dec                 & Radius & $V_\text{exp}$ & Dynamical & Shell mass             & Kinetic                 & Cluster mass,              & Notes  \\
ID     &         & (deg)                & (deg)               & (pc)   & (km~s$^{-1}$)  & age (Myr) & (M$_\odot$)            & energy (erg)            & $M_\text{cl}$  (M$_\odot$) &         \\
\hline
1      & NGC~0628 & $\ang{24;09;10.4}$  & $\ang{15;46;33.3}$  & $105$  & $10.5$         & $2.4$    & $15.32 \times 10^{7}$  & $16.00 \times 10^{51}$  & $20.73 \times 10^{4}$      & Arm \\
13     & NGC~1087 & $\ang{41;36;33.3}$  & $\ang{-0;29;25.1}$  & $105$  & $14.5$         & $1.8$    & $16.27 \times 10^{7}$  & $23.47 \times 10^{51}$  & $41.97 \times 10^{4}$      & None\\
16     & NGC~1300 & $\ang{49;53;53.9}$  & $\ang{-19;24;17.1}$ & $149$  & $10.7$         & $3.4$    & $27.74 \times 10^{7}$  & $29.40 \times 10^{51}$  & $27.33 \times 10^{4}$      & Arm and inter-arm\\
18     & NGC~1365 & $\ang{53;25;51.7}$  & $\ang{-36;08;49.6}$ & $144$  & $25.5$         & $1.4$    & $111.61 \times 10^{7}$ & $283.18 \times 10^{51}$ & $649.88 \times 10^{4}$     & Arm\\
22     & NGC~1385 & $\ang{54;22;25.5}$  & $\ang{-24;29;50.8}$ & $167$  & $9.8$          & $4.2$    & $60.66 \times 10^{7}$  & $59.15 \times 10^{51}$  & $45.07 \times 10^{4}$      & Inter-arm\\
23     & NGC~1433 & $\ang{55;28;25.4}$  & $\ang{-47;13;7.1}$  & $131$  & $7.9$          & $4.0$    & $2.33 \times 10^{7}$   & $1.83 \times 10^{51}$   & $1.43 \times 10^{4}$       & None \\
25     & NGC~1512 & $\ang{60;57;40.7}$  & $\ang{-43;21;51.3}$ & $188$  & $8.5$          & $5.4$    & $4.79 \times 10^{7}$   & $4.05 \times 10^{51}$   & $2.38 \times 10^{4}$       & Bar and inter-arm\\
28     & NGC~1566 & $\ang{64;59;4.8}$   & $\ang{-54;57;0.7}$  & $88$   & $6.9$          & $3.1$    & $1.75 \times 10^{7}$   & $1.20 \times 10^{51}$   & $1.22 \times 10^{4}$       & Inter\\
40     & NGC~1627 & $\ang{71;23;19.6}$  & $\ang{-59;15;5.2}$  & $171$  & $11.0$         & $3.8$    & $18.39 \times 10^{7}$  & $20.13 \times 10^{51}$  & $16.80 \times 10^{4}$      & Inter-arm\\
43     & NGC~2835 & $\ang{139;27;24.3}$ & $\ang{-22;20;32.1}$ & $60$   & $5.6$          & $2.6$    & $0.73 \times 10^{7}$   & $0.41 \times 10^{51}$   & $0.50 \times 10^{4}$       & Arm and inter-arm\\
45     & NGC~3351 & $\ang{161;00;9.4}$  & $\ang{11;42;35.2}$  & $50$   & $6.3$          & $2.0$    & $0.42 \times 10^{7}$   & $0.27 \times 10^{51}$   & $0.43 \times 10^{4}$       & Disc\\
47     & NGC~3627 & $\ang{170;02;56.2}$ & $\ang{12;58;41.5}$  & $100$  & $15.0$         & $1.6$    & $29.49 \times 10^{7}$  & $44.01 \times 10^{51}$  & $85.25 \times 10^{4}$      & Arm\\
57     & NGC~4252 & $\ang{184;41;26.6}$ & $\ang{14;24;7.6}$   & $167$  & $11.8$         & $3.4$    & $27.50 \times 10^{7}$  & $32.29 \times 10^{51}$  & $29.63 \times 10^{4}$      & Arm\\
70     & NGC~4303 & $\ang{185;27;29.5}$ & $\ang{4;28;46.0}$   & $141$  & $8.5$          & $4.0$    & $17.04 \times 10^{7}$  & $14.42 \times 10^{51}$  & $11.30 \times 10^{4}$      & Arm\\
79     & NGC~4321 & $\ang{185;44;29.6}$ & $\ang{15;49;48.9}$  & $116$  & $11.9$         & $2.4$    & $3.14 \times 10^{7}$   & $3.72 \times 10^{51}$   & $4.95 \times 10^{4}$       & Arm\\
81     & NGC~4535 & $\ang{188;34;16.6}$ & $\ang{8;12;32.0}$   & $299$  & $22.0$         & $3.3$    & $130.99 \times 10^{7}$ & $286.73 \times 10^{51}$ & $273.65 \times 10^{4}$     & Arm\\
82     & NGC~5068 & $\ang{199;42;30.1}$ & $\ang{-21;02;17.9}$ & $51$   & $9.0$          & $1.4$    & $2.23 \times 10^{7}$   & $1.99 \times 10^{51}$   & $4.58 \times 10^{4}$       & None\\
85     & NGC~7496 & $\ang{347;26;6.3}$  & $\ang{-43;25;25.4}$ & $136$  & $6.9$          & $4.8$    & $2.36 \times 10^{7}$   & $1.62 \times 10^{51}$   & $1.06 \times 10^{4}$       & Disc and none\\

\end{tabular}
    }
    \label{tab:props}
\end{table*}

\begin{table*}[]
\centering 
    \caption{Mass averaged weighted ages, total mass, and number of clusters found for the multi-scale association catalogues found inside the superbubbles. Only the first superbubble from each galaxy is listed.}
\resizebox{\textwidth}{!}{
\begin{tabular}{rrrrrrrrrr}
Bubble & Age          & Mass                          & No.   & Age          & Mass                          & No.   & Age           & Mass                          & No.  \\
ID     & 16~pc (Myr)  & 16~pc (\msun)                 & 16~pc & 32~pc (Myr)  & 32~pc (\msun)                 & 32~pc & 64~pc (Myr)   & 64~pc (\msun)                 & 64~pc \\
\hline
$1$    & $1.2\pm1.1$  & $18.92\pm3.12 \times 10^{4}$  & $8$   & $1.0\pm1.0$  & $22.71\pm5.60 \times 10^{4}$  & $2$   & $1.0\pm0.5$   & $21.28\pm3.28 \times 10^{4}$  & $1$\\
$13$   & $2.0\pm0.6$  & $10.45\pm3.49 \times 10^{4}$  & $6$   & $1.6\pm0.6$  & $16.71\pm5.14 \times 10^{4}$  & $3$   & $1.8\pm2.2$   & $14.75\pm16.03 \times 10^{4}$ & $2$\\
$16$   & $4.8\pm1.0$  & $0.77\pm0.24 \times 10^{4}$   & $2$   & $4.0\pm0.6$  & $7.23\pm9.38 \times 10^{4}$   & $4$   & $6.4\pm0.9$   & $10.27\pm4.30 \times 10^{4}$  & $2$\\
$18$   & $4.7\pm1.7$  & $13.46\pm20.04 \times 10^{4}$ & $6$   & $2.2\pm4.6$  & $63.10\pm40.20 \times 10^{4}$ & $6$   & $7.3\pm3.1$   & $19.24\pm51.79 \times 10^{4}$ & $4$\\
$22$   & $4.0\pm0.7$  & $11.2\pm6.27 \times 10^{4}$   & $7$   & $4.1\pm1.0$  & $26.02\pm21.96 \times 10^{4}$ & $8$   & $7.0\pm1.0$   & $19.34\pm1.23 \times 10^{4}$  & $3$\\
$23$   & $1.1\pm0.9$  & $0.61\pm0.18 \times 10^{4}$   & $3$   & $1.3\pm0.9$  & $0.71\pm0.21 \times 10^{4}$   & $3$   & $7.0\pm0.5$   & $0.13\pm0.02 \times 10^{4}$   & $1$\\
$25$   & $3.8\pm1.1$  & $1.17\pm0.86 \times 10^{4}$   & $5$   & $3.7\pm0.8$  & $1.40\pm1.13 \times 10^{4}$   & $3$   & $6.0\pm5.0$   & $0.90\pm7.34 \times 10^{4}$   & $1$\\
$28$   & $7.2\pm1.0$  & $0.68\pm0.12 \times 10^{4}$   & $2$   & $5.5\pm1.8$  & $1.06\pm3.22 \times 10^{4}$   & $3$   & $1.0\pm6.0$   & $4.37\pm6.34 \times 10^{4}$   & $1$\\
$40$   & $7.8\pm0.6$  & $1.91\pm0.23 \times 10^{4}$   & $3$   & $2.2\pm1.1$  & $15.57\pm13.78 \times 10^{4}$ & $6$   & $1.0\pm2.7$   & $64.04\pm47.63 \times 10^{4}$ & $4$\\
$43$   & $2.0\pm0.5$  & $0.26\pm0.06 \times 10^{4}$   & $1$   & $2.4\pm0.6$  & $0.64\pm0.25 \times 10^{4}$   & $2$   & $3.0\pm1.0$   & $1.21\pm1.48 \times 10^{4}$   & $1$\\
$45$   & $-$          & $-$                           & $0$   & $-$          & $-$                           & $0$   & $10.0\pm0.5$  & $0.24\pm0.04 \times 10^{4}$   & $1$\\
$47$   & $4.4\pm0.7 $ & $5.07\pm10.71 \times 10^{4}$  & $3$   & $4.0\pm1.4$  & $7.51\pm14.63 \times 10^{4}$  & $2$   & $7.0\pm25.0$  & $2.17\pm14.77 \times 10^{4}$  & $1$\\
$57$   & $5.3\pm1.2$  & $10.47\pm9.76 \times 10^{4}$  & $8$   & $6.8\pm2.6$  & $13.07\pm20.67 \times 10^{4}$ & $5$   & $6.8\pm1.0$   & $13.81\pm5.05 \times 10^{4}$  & $3$\\
$70$   & $5.4\pm1.3$  & $12.02\pm4.93 \times 10^{4}$  & $5$   & $6.8\pm0.9$  & $9.22\pm0.67 \times 10^{4}$   & $3$   & $7.0\pm33.0$  & $8.57\pm18.37 \times 10^{4}$  & $1$\\
$79$   & $1.0\pm1.4$  & $8.27\pm1.66 \times 10^{4}$   & $2$   & $5.0\pm0.6$  & $4.63\pm1.54 \times 10^{4}$   & $4$   & $5.9\pm0.9$   & $2.86\pm4.98 \times 10^{4}$   & $3$\\
$81$   & $3.5\pm0.7$  & $21.22\pm6.67 \times 10^{4}$  & $21$  & $3.4\pm1.2$  & $38.98\pm18.06 \times 10^{4}$ & $14$  & $3.2\pm1.2$   & $42.42\pm21.16 \times 10^{4}$ & $9$\\
$82$   & $-$          & $-$                           & $0$   & $11.0\pm2.0$ & $0.17\pm0.08 \times 10^{4}$   & $1$   & $11.0\pm15.0$ & $0.15\pm0.33 \times 10^{4}$   & $1$\\
$85$   & $7.6\pm0.6$  & $2.25\pm2.69 \times 10^{4}$   & $2$   & $8.0\pm0.5$  & $3.75\pm0.42 \times 10^{4}$   & $2$   & $7.8\pm1.9$   & $4.50\pm6.87 \times 10^{4}$   & $2$\\

\end{tabular}
    }
    \label{tab:hst_props}
\end{table*}

\begin{acknowledgements}
We would like to thank the referee for their constructive feedback that helped improve the quality of this work. This work was carried out as part of the PHANGS collaboration. Based on observations collected at the European Southern Observatory under ESO programmes 094.C-0623 (PI: Kreckel), 095.C-0473, 098.C-0484 (PI: Blanc), 1100.B-0651 (PHANGS-MUSE; PI: Schinnerer), as well as 094.B-0321 (MAGNUM; PI: Marconi), 099.B-0242, 0100.B-0116, 098.B-0551 (MAD; PI: Carollo) and 097.B-0640 (TIMER; PI: Gadotti). Furthermore, this paper makes use of the following ALMA data: \linebreak
ADS/JAO.ALMA\#2012.1.00650.S, \linebreak
ADS/JAO.ALMA\#2013.1.01161.S, \linebreak
ADS/JAO.ALMA\#2015.1.00925.S, \linebreak
ADS/JAO.ALMA\#2015.1.00956.S, \linebreak
ADS/JAO.ALMA\#2017.1.00392.S, \linebreak
ADS/JAO.ALMA\#2017.1.00886.L, \linebreak
ADS/JAO.ALMA\#2018.1.01651.S. \linebreak
ALMA is a partnership of ESO (representing its member states), NSF (USA) and NINS (Japan), together with NRC (Canada), MOST and ASIAA (Taiwan), and KASI (Republic of Korea), in cooperation with the Republic of Chile. The Joint ALMA Observatory is operated by ESO, AUI/NRAO and NAOJ. In addition, this research uses observations made with the NASA/ESA Hubble Space Telescope obtained from the Space Telescope Science Institute, which is operated by the Association of Universities for Research in Astronomy, Inc., under NASA contract NAS 5–26555. These observations are associated with program 15654 and can be accessed at: \url{https://dx.doi.org/10.17909/t9-r08f-dq31}. 
Distances for the galaxies used in this work were measured in \cite{freedman_measurements_2021}, \cite{nugent_toward_2006}, \cite{jacobs_extragalactic_2009}, \cite{kourkchi_galaxy_2017}, \cite{shaya_action_2017}, \cite{kourkchi_cosmicflows-3_2020}, \cite{anand_distances_2021}, and \cite{scheuermann_planetary_2022}.
EJW, KK, OE, JEM-D and FS gratefully acknowledge funding from the German Research Foundation (DFG) in the form of an Emmy Noether Research Group (grant number KR4598/2-1, PI Kreckel). MC gratefully acknowledges funding from the Deutsche Forschungsgemeinschaft (DFG) through an Emmy Noether Research Group, grant number CH2137/1-1. JMDK gratefully acknowledges funding from the European Research Council (ERC) under the European Union’s Horizon 2020 research and innovation programme via the ERC Starting Grant MUSTANG (grant agreement number 714907). COOL Research DAO is a Decentralised Autonomous Organisation supporting research in astrophysics aimed at uncovering our cosmic origins. ES acknowledges funding from the European Research Council (ERC) under the European Union’s Horizon 2020 research and innovation programme (grant agreement No. 694343). IP acknowledges funding by the European Research Council through ERC-AdG SPECMAP-CGM, GA 101020943. ATB would like to acknowledge funding from the European Research Council (ERC) under the European Union’s Horizon 2020 research and innovation programme (grant agreement No.726384/Empire). PSB acknowledges funding from the Spanish Ministry of Science and Innovation, under the grant PID2019-107427-GB-C31. SCOG and RSK acknowledge financial support from the DFG via the collaborative research center (SFB 881, Project-ID 138713538) “The Milky Way System” (subprojects A1, B1, B2 and B8), from the Heidelberg Cluster of Excellence “STRUCTURES” in the framework of Germany’s Excellence Strategy (grant EXC-2181/1, Project-ID 390900948) and from the European Research Council (ERC) via the ERC Synergy Grant “ECOGAL” (grant 855130). MB acknowledges support from FONDECYT regular grant 1211000 and by the ANID BASAL project FB210003. KG is supported by the Australian Research Council through the Discovery Early Career Researcher Award (DECRA) Fellowship (project number DE220100766) funded by the Australian Government. KG is supported by the Australian Research Council Centre of Excellence for All Sky Astrophysics in 3 Dimensions (ASTRO~3D), through project number CE170100013. 
\end{acknowledgements}

%
   \bibliographystyle{aa} 
   \bibliography{paper} 
%

\end{document}